\documentclass[aps,10pt,prd,groupedaddress,nofootinbib,preprint,eqsecnum]{revtex4-1}
\usepackage[utf8]{inputenc}
\usepackage{hyperref}
\usepackage{xcolor}
\usepackage{graphicx}
\usepackage{amsmath,amssymb}
\usepackage{bm}
\newcommand{\pa}{\partial}

\newcommand{\dph}{\delta\phi}

\newenvironment{equations}{\equation\aligned}{\endaligned\endequation}

\def\beq{\begin{equation}}
	\def\eeq{\end{equation}}
\def\ba{\begin{equations}}
	\def\ea{\end{equations}}
\def\bc{\begin{center}}
	\def\ec{\end{center}}

\newcommand{\dt}[1]{\frac{d}{dt}\left(#1\right)}

\def\pip{\pi_\phi}
\def\pia{\pi_\alpha}
\def\vp{\varphi}

\def\mn{{\mu\nu}}
\def\ij{{ij}}

\def\ch{{\cal H}}

\def\cl{{\cal L}}
\def\cR{{\cal R}}

\def\pa{\partial}

\def\e{{\rm e}}
\def\dphi{\delta\phi}
\newcommand{\pb}[2]{\left\{{#1},{#2}\right\}}
\def\lie{\raisebox{0.2ex}{\textbf{--}}\hspace{-2.5mm}{\cl}}

\begin{document}
	\preprint{YITP-17-104}
	\title{\boldmath Hamiltonian approach to 2nd order gauge invariant cosmological perturbations}
	
	\author{\textsc{Guillem Dom\`enech$^{a,b}$}}
	\email{guillem.domenech{}@{}yukawa.kyoto-u.ac.jp}
	
	\author{\textsc{Misao Sasaki$^{a,b}$} }
	\email{misao{}@{}yukawa.kyoto-u.ac.jp}

	\affiliation{
		$^{a}$\small{Center for Gravitational Physics, Yukawa Institute for Theoretical Physics, Kyoto University,
			Kyoto 606-8502, Japan}\\\\
		$^{a}$\small{International Research Unit of Advanced Future Studies, Kyoto University, Japan}
	}
	
	\begin{abstract} 
In view of growing interest in tensor modes and their possible detection, we clarify
the definition of tensor modes up to 2nd order in perturbation theory within 
the Hamiltonian formalism. Like in gauge theory, in cosmology the Hamiltonian 
is a suitable and consistent approach to reduce the gauge degrees of freedom. 
In this paper we employ the Faddeev-Jackiw method of Hamiltonian reduction.
An appropriate set of gauge invariant variables that describe the dynamical 
degrees of freedom may be obtained by suitable canonical transformations
in the phase space. 
We derive a set of gauge invariant variables up to 2nd order in 
perturbation expansion and for the first time we reduce the 3rd order action 
without adding gauge fixing terms. 
In particular, we are able to show the relation between the
uniform-$\phi$ and Newtonian slicings,
and study the difference in the definition of tensor modes in these two
slicings.
	\end{abstract}
	
	\date{\today}
	\maketitle
	
	\section{\label{sec:intro} Introduction}

	 The detection of gravitational waves \cite{Abbott:2016blz} has 
	 opened up the era of gravitational wave physics/astronomy, and
	 given us more hope of detecting gravitational waves (GWs) generated 
	 during and/or after inflation. GWs generated during inflation
	 (more appropriately called the tensor modes) could be seen in the Cosmic 
	 Microwave Background (CMB) if the resulting B-modes would be detected 
	 (see Ref.~\cite{Kamionkowski:2015yta,Guzzetti:2016mkm} and references therein). 
	 The possibility to detect their effects in the temperature fluctuations has
	 been discussed as well. 
	 For example, a model in which the tensor modes are significantly mixed 
	 with the scalar modes could explain a scale dependence in the temperature 
	 bispectrum \cite{Domenech:2017kno}. Axion-like and gauge spectator fields 
	 could enhance the amplitude of the tensor modes and render them 
	 chiral  \cite{Sorbo:2011rz,Biagetti:2013kwa,Agrawal:2017awz,Fujita:2017jwq}. 
	 Parametric resonances in massive gravity may also enhance tensor modes 
	 during reheating \cite{Lin:2015nda}. After the end of inflation, 
	 the 2nd order scalar perturbations source GWs which may be 
	 detectable~\cite{Ananda:2006af,Baumann:2007zm,Alabidi:2012ex}, etc.
	 
	 In this situation, and especially when higher order effects are concerned, 
	 it is fundamentally important to distinguish tensor modes from scalar modes 
	 in a given gauge, since the definition of the tensor perturbation
	 depends on the choice of gauge. 
	 The decomposition theorem shows us that at linear order scalar, vector 
	 and tensor modes decouple from each other,
	 and hence we can separately build gauge invariant 
	 variables \cite{Bardeen:1980kt,Kodama:1985bj,Mukhanov:1990me}. 
	 At 2nd order in perturbation, however, the situation becomes more involved 
	 as the decomposition theorem does not apply any more;
	 scalar, vector and tensor modes mix with each other. 
	 There have been several works in this direction, e.g. 
	 Refs.~\cite{Bruni:1996im,Nakamura:2004rm,Langlois:2005ii,Finelli:2006wk,Nakamura:2006rk,Malik:2008im,Naruko:2013aaa,Carrilho:2015cma,Bertacca:2015mca}. 
	 The situation is better understood only when one focuses on scalar modes.
	 For example, there exists a conservation law for the non-linear 
	 curvature perturbation on super-horizon scales \cite{Lyth:2004gb,Naruko:2011zk}. 
	 More generally, gauge invariant variables may be built by computing 
	 the Lie derivatives of the metric and matter fields and then 
	 finding gauge invariant combinations of them. 
	 This is appropriate as long as one wants to compare quantities in different gauges 
	 but it is not suitable for finding a set of gauge invariant variables 
	 which represent the dynamical degrees of freedom of the system. 
	 The Hamiltonian approach is most suited for this purpose, much like the case
	 of gauge theory.
	 In this direction, there is work by Langlois \cite{Langlois:1994ec} that deals
	 with gauge issues at 1st order perturbation theory. 
	 Recently, Ref.~\cite{Nandi:2015ogk} studied the non-linear Hamiltonian 
	 with the gauge fixed. 
	 
	 We dedicate this work to study 2nd order gauge invariant cosmological 
	 perturbations in the Hamiltonian formalism. Like in a gauge theory, 
	 the Hamiltonian provides 
	 insights into the structure and symmetries of the theory as well as 
	 the number of dynamical degrees of freedom.
	 It does not only constitute a complementary approach to the already existing results in the literature but also clarifies in a concise manner the definitions
	 of correct gauge invariant variables and, in particular, the mixing between 
	 scalar and tensor modes. Here we employ the Faddeev-Jackiw method of
	 Hamiltonian reduction~\cite{Faddeev:1988qp}.
	 The main advantage is that the Hamiltonian in general
	 relativity is the generator of infinitesimal coordinate transformations by itself
	 and, therefore, provides a self-consistent way to reduce 
	 the degrees of freedom of our system.
	 In contrast to the Lagrangian approach, the reduction of the 3rd order action 
	 will be given by a suitable canonical transformation in the phase space. 
	 For simplicity, we will consider a canonical scalar field but the 
	 generalization to non-canonical fields is 
	 straightforward.~\footnote{A drawback of the Hamiltonian approach is 
	 	that one needs to know the specific form of the kinetic term for the scalar 
	 	field, e.g. in general K-inflation theory. Nevertheless, the gauge invariant 
	 	variables thus obtained may apply to any theory as the symplectic structure 
	 	still holds. Although the definition of the canonical momenta might differ.}
	 In doing so, we succeeded in reducing for the first time the 3rd order action 
	 without adding any gauge fixing term.
	 We will first work with the variables that coincide with perturbations 
	 in the uniform-$\phi$ slicing. After obtaining all the relevant equations,
	 we then derive the transformation rules to the Newtonian slicing.
	 
	 This paper is organized as follows. In section \ref{sec:1}, we compute the Hamiltonian in the conformal decomposition without any perturbative expansion. Meanwhile, we review the Poisson algebra and the fact that the Hamiltonian is the generator of coordinate transformations. In section \ref{sec:2.5}, we expand the Hamiltonian and explain the method to find the gauge invariant variables and the reduction of the action. In section \ref{sec:3}, we apply it to 1st order in perturbation expansion reviewing the results of Ref.\cite{Langlois:1994ec} for the 2nd order Hamiltonian. Then in section \ref{sec:4}, we study the 2nd order perturbation theory. We start by computing the gauge invariant variables which coincide with the uniform density slicing and then we proceed to the reduction of the 3rd order Hamiltonian without any gauge fixing. We dedicate section \ref{sec:newtonian} to show how to go from the uniform-$\phi$ to Newtonian slicings up to 2nd order. We compare the resulting equations of motion with existing results in the literature. Finally in section \ref{sec:5}, we summarize our work and discuss future directions. 
	 Detailed derivations and expressions of the equations in the text
	 are provided in the Appendices.
	
	
	\section{Hamiltonian in the conformal decomposition\label{sec:1}}
	Let us briefly review the Hamiltonian approach to general relativity. For our purposes, it is convenient to work in the ADM formalism \cite{Arnowitt:1962hi} with the conformal decomposition of the spatial metric. In this case, the 4D line element is given by
	\ba\label{eq:confdecom}
	ds^2&=g_{\mn}dx^\mu dx^\nu
	=-N^2dt^2+{\rm e}^{2\Psi}\Upsilon_{ij}\left(dx^i+N^idt\right)\left(dx^j+N^jdt\right)\,,
	\ea
	where $g_{\mn}$ is the metric of our space-time, $N$ is the lapse function, $N^i$ is the shift vector, $i=1,\,2,\,3$ and we have chosen that
	\ba\label{eq:derdet}
	\frac{\pa}{\pa t}\det \Upsilon =\Upsilon^{ij}\frac{\pa}{\pa t}\Upsilon_{ij}=0
	\ea
	and therefore the dynamical degrees of freedom of the metric's determinant are encoded in $\Psi$ and $\Upsilon$ contains the traceless degrees of freedom. We will see later that the perturbation of $\Psi$ corresponds to the non-linearly conserved comoving curvature perturbation on super-horizon scales. The perturbation of $\Upsilon_{ij}$ contains the non-linearly conserved tensor perturbations on the uniform-$\phi$ slicing on super-horizon scales.
	
	For simplicity, we consider a self-gravitating canonical scalar field $\Theta$ whose action is given by
	\ba\label{eq:action1}
		S= \int d^4x \sqrt{-g} \left\{\frac{1}{2}R-\frac{1}{2}g^{\mu\nu}\pa_\mu\Theta\pa_\nu\Theta-V(\Theta)\right\}\,,
	\ea
	where $g$ is the determinant of $g_{\mn}$, $R$ is the 4D Ricci scalar, $\pa_\mu\equiv\pa/\pa x^\mu$ and $V(\Theta)$ is a general potential for the scalar field $\Theta$.\footnote{The generalization to a given K-inflation or Horndeski theory is involved but straightforward.}
	
    In the conformal decomposition Eq.~\eqref{eq:confdecom}, the action reads
	\ba
		\begin{split}
			S=\int d^3x dt N \bigg\{ {\rm e}^{\Psi}&\Bigg(\frac{1}{2}R^{(3)}[\Upsilon]-2D^iD_i\Psi-D^i \Psi D_i \Psi-\frac{1}{2}(D\Theta)^2\Bigg)\\&
			+\frac{1}{2}{\rm e}^{3\Psi}\left(E_{ij}E^{ij}-\frac{2}{3}K^2+(\lie_n\Theta)^2\right)-{\rm e}^{3\Psi}V(\Theta)\bigg\}\,,
		\end{split}
	\ea

	where $R^{(3)}[\Upsilon]$ and $D_i$ are respectively the 3D Ricci scalar and the covariant derivative corresponding to $\Upsilon_{ij}$. Note that we used that $\det\Upsilon=1$ since we will work in a flat FLRW metric in Cartesian coordinates.\footnote{It could be generalized to any spatial metric which satisfies $\pa_t\left(\det\Upsilon\right)=0$. We set $\det\Upsilon = 1$ for simplicity.} We denoted as $\lie_n\equiv n^\mu\pa_\mu$ the Lie derivative along the hyper-surface orthonormal direction $n^\mu$, where $n_\mu dx^\mu=-Ndt$. 

The extrinsic curvature corresponding to $\Psi$ and $\Upsilon_{ij}$ is respectively given by
	\ba
		K=3\lie_n{\Psi}-\frac{1}{N}D_{k}N^{k}
	\ea
	and
	\ba
		E_{ij}=\frac{1}{2N}\left(\dot{\Upsilon}_{ij}-2D_{(i}N_{j)}+\frac{2}{3}\Upsilon_{ij}D_{k}N^{k}\right)\,,
	\ea
	where $\dot Q_A\equiv\pa_t Q_A$ with $Q_A=\left\{\Theta,\Psi,\Upsilon_{ij}\right\}$ and it should be noted that $\Upsilon^{ij}E_{ij}=0$ by using Eq.~\eqref{eq:derdet}. With these definitions we proceed to the computation of the Hamiltonian as usual. We define the conjugate momenta as $\Pi_A\equiv{\delta {\cal L}}/{\delta \dot{Q}_A}$ where $\cl$ is the Lagrangian density, i.e. $S=\int dt \,d^3x \,\cl$. In this way, we have that
	\ba
    \Pi_{\Theta}= {\rm e}^{3\Psi}\lie_n\Theta\quad;\quad \Pi_{\Psi}=-2 {\rm e}^{3\Psi}\,K\quad;\quad	\Pi^{ij}&=\frac{ {\rm e}^{3\Psi}}{2} E^{ij}
	\ea
	and the Hamiltonian density is given by
	\ba
		\ch=\Pi^{ij}\dot\Upsilon_{ij}+\Pi_\Psi\dot\Psi+\Pi_\Theta\dot\Theta-{\cal L}= N\ch_N+N^i\ch_i\,,
	\ea
	where
	\ba
		\begin{split}
			\ch_N=&{{\rm e}^{-3\Psi}}\left(2\Pi^{ij}\Pi_{ij}-\frac{\Pi_\Psi^2}{12}+\frac{\Pi_{\Theta}^2}{2}\right)+{\rm e}^{3\Psi}V(\Theta)\\&
			+ {\rm e}^{\Psi}\bigg(2\Upsilon^{ij}D_iD_j\Psi+\Upsilon^{ij}D_i \Psi D_j \Psi-\frac{1}{2}R^{(3)}+\frac{1}{2}\Upsilon^{ij} D_i \Theta D_j \Theta\bigg)
		\end{split}
	\ea
	and
	\ba
		\begin{split}
			\ch_i&=\Pi_\Theta D_i\Theta+\Pi_\Psi D_i\Psi-\frac{1}{3}D_i\Pi_\Psi-2\Upsilon_{ij}D_k\Pi^{kj}\,.
		\end{split}
	\ea
	As usual the Lapse function $N$ and the shift vector $N^i$ act as a Lagrange multipliers since $\Pi_N=\Pi_{N^i}=0$. The variation with respect to them yields the so-called Hamiltonian and Momentum constraint, i.e. $\ch_N=\ch_i=0$. In a general gauge, all these four constraints are first class (for a review on the Hamiltonian formalism see \cite{Bojowald:2010qpa}). This means that our system, which initially contained $11$ degrees of freedom ($10$ for the metric and $1$ from the scalar field), contains only $3$ dynamical degrees of freedom, $1$ scalar and $2$ tensor modes. Thus, to study cosmological perturbations up to 2nd order we will have to solve the constraints and reduce the number of degrees of freedom of our Hamiltonian.

	\subsection{Poisson algebra and coordinate transformations}
	It is important to note that, in the Hamiltonian formalism, the phase space generates a symplectic manifold,\footnote{Using the phase space we can build a non-degenerate closed 2-form by taking the exterior derivative between $dQ_A$ and $dP^A$. This 2-form is the base for the symplectic structure, see Chap.~8 of Ref.~\cite{vogtmann1997mathematical} for more details.} In this symplectic geometry, one can show that the Lie derivative of a quantity along a Hamiltonian vector field is equal to the Poisson bracket of that quantity with the generator of the Hamiltonian vector field \cite{vogtmann1997mathematical}. For our interest, the Lie derivative along the Hamiltonian flow, i.e. along trajectories in the phase space, is equal to the Poisson bracket with the Hamiltonian. In particular, the Lie derivative of a quantity, say $f$, along a direction $\epsilon^\mu$ is given by
		\ba
		\lie_{{\epsilon}}f=\pb{f}{\epsilon^\mu\ch_\mu}\,,
		\ea
	where $\ch_\mu=\left\{\ch_N,\ch_i\right\}$ and the Poisson bracket between two functions $f$ and $g$ is given by
\ba
\pb{{f}}{g}=\frac{\delta {f}}{\delta{Q_A}}\frac{\delta {g}}{\delta \Pi^{A}}-\frac{\delta {f}}{\delta\Pi^{A}}\frac{\delta {g}}{\delta {Q_A}}\,.
\ea
Basically, we are following the change of a given variable in the phase space with a change of coordinates along $\epsilon^\mu$. Thus, the Hamiltonian not only describes the time evolution of the system but it is also the generator of coordinate transformations.\footnote{In other words, a first class constraint is the generator of a symmetry of the Lagrangian and the variables transform according to the poisson bracket with the constraint. In our present study, the first class constraints are the Hamiltonian and Momentum constraints and the symmetry is diffeomorphism invariance.}  For a general variable $Q_A$ we can write its change under an infinitesimal coordinate transformation as
	\ba
	Q'_A=Q_A+\lie_{\epsilon}Q_A=Q_A+\pb{Q_A}{\epsilon^\mu \ch_\mu}\,.
	\ea
Now, we can check that the Hamiltonian and Momentum constraints generate a closed Poisson algebra. Concretely, for the smeared form of the constraints, i.e. $H[N]\equiv\int d^3x N \ch$ and $H_i[N^i]\equiv\int d^3x N^i \ch_i$, we have
	\ba
	\left\{H[N],H[M]\right\}&=\int d^3x \,\e^{-2\Psi}\left(ND^iM-MD^iN\right)\ch_i\,,\\
	\left\{H_i[N^i],H[M]\right\}&=\int d^3x \,N^kD_kM\,\ch\,,\\
	\left\{H_i[N^i],H_i[M^i]\right\}&=\int d^3x \,\left(N^kD_kM^l-M^kD_kN^l\right)\,\ch_l\,.
	\ea

	This will be our starting point to compute the gauge invariant variables at 2nd order and the corresponding reduced Hamiltonian.
	
	\section{Perturbed Hamiltonian\label{sec:2.5}}
	Let us now study cosmological perturbations during inflation within the Hamiltonian formalism up to 2nd order. For simplicity, we focus on a spatially flat FLRW	background in Cartesian coordinates. The latter choice of coordinates simplifies later calculations since in that case $\det \Upsilon=1$ and indexes are raised and lowered by $\delta_{ij}$. In addition, we will neglect vector modes since in single field inflation they rapidly decay \cite{Kodama:1985bj,Mukhanov:1990me}. We expand into a time dependent background and perturbations as follows. The conformal degree of freedom, the inflaton and the lapse and shift are split as
	\ba
		\Psi=\alpha(t)+\psi(t,\mathbf{x}) \quad {\rm ,}&\quad \Theta=\phi(t)+\varphi(t,\mathbf{x})\,,
		\\
				\Pi_\Psi=\pi_{\alpha}(t)+\pi_\psi(t,\mathbf{x})\quad{\rm ,}&\quad \Pi_\Theta=\pi_\phi(t)+\pi_\varphi(t,\mathbf{x})\,,\\
				N=1+A \quad&{\rm and}\quad N_i=a^{-2}\pa_i B\,,
	\ea
	where $\alpha\equiv\ln a$, $\pi_\alpha=-6a^3H$, $ \pi_\phi=a^3\dot\phi $, $a$ is the scale factor of the FLRW expanding universe with Hubble rate $H\equiv{\dot a}/{a}$ and $\phi$ is the background value of the inflaton. We keep the notation $\pia$ and $\pip$ for computational convenience, as we will see later. At the end of the paper we will present the formulas in the usual convention. The perturbations of the spatial metric are given by
	\ba
		Y_{ij}\equiv\left[\ln\Upsilon\right]_{ij}=\gamma_{ij}+2D_{ij}E
	\ea
	where $D_{ij}\equiv\partial_i\partial_j - \frac{1}{3}\delta_{ij}\Delta$, $\Delta\equiv\pa_i\pa^i$ and one should understand $E$ and $\gamma_{ij}$ respectively as the non-transverse and transverse modes, namely
		\ba
		E=\frac{3}{4}\Delta^{-2}\pa_k\pa_lY_{kl}\quad{\rm and}\quad \gamma_{ij}=\widehat{TT}_{ij}\,^{ab}Y_{ab}\,,
		\ea
		where $\Delta^{-1}$ is the inverse Laplacian operator which we assume to be well defined and that the fields vanish at infinity. $\widehat{TT}_{ij}\,^{ab}$ is the transverse-traceless projector given in Eq.~\eqref{eq:tt} in App.\ref{app:perturbvar}. Last, the canonical momenta is expanded as
		\ba
		\Pi^{ij}=P^{l(i}\delta_{lk}\Upsilon^{j)k} + O(4)
		\quad
		{\rm with}
		\quad
	P_{ij}=\pi_{ij}+\frac{3}{4}D_{ij}\Delta^{-2}\pi_E\,,
		\ea
		where $\pi_E=2\pa_a\pa_bP^{ab}$ and $\pi_{ij}=\widehat{TT}^{ij}\,_{ab}P^{ab}$. Note that the expansion of $\Pi^{ij}$ is only valid up to 4th order. We are now ready to expand the Hamiltonian, study the gauge transformations and find the gauge invariant variables. Before going into details let us show the meaning of each momenta in the usual variables up to 1st order. We have
	\ba
	\pi_\psi =&-6a^3\left(\dot\psi+H\left(3\psi-A\right)-\frac{1}{3}\Delta B\right)\,,\\
	\pi_\vp=&a^3\left(\dot\vp+\dot\phi\left(3\psi-A\right)\right)\,,\\
	\pi_E=&\frac{2}{3}\Delta\Delta\left(\dot E -B\right)\quad{\rm and}\quad
	\pi_{ij}=\frac{1}{4}a^3\dot\gamma_{ij}\,.
	\ea
	In particular note that $\Delta^{-2}\pi_E$ is proportional to the shear, i.e. $\sigma_g\equiv \dot E-B$, and that $\pi_\psi$ is proportional to the expansion parameter.

 Before going into details, let us note that the reduction of the Hamiltonian will essentially be equivalent to a suitable time dependent canonical transformation. Due to the time dependence, the Hamiltonian will receive correction terms in addition to the mere change of variables. This is clearer in the Lagrangian formalism where a coordinate transformation does not modify the Lagrangian except for a total time derivative. Formally we can write the change of the Lagrangian from a set of canonical variables $\left\{q_a,p^a\right\}$ to $q_a\to q_a+\delta q_a$ and $p^a\to p^a+\delta p^a$ as\footnote{Note that here $\delta$ refers to change in the canonical variables, not an infinitesimal displacement.}
\ba
\delta \cl=\delta \left(p^a \dot q_a \right)-\delta\ch=\rm{(total\,\,derivative)}\,,
\ea
where in our case $q_a=\left\{\psi,\vp,E,\gamma_{ij}\right\}$ and $p^a=\left\{\pi_\psi,\pi_\vp,\pi_E,\pi_{ij}\right\}$. The Hamiltonian changes as $\ch\to\ch+\delta\ch$. In this way it is clear that if the canonical transformation has a time dependence then $\delta\ch\neq0$ in general. We will find the change in the Hamiltonian by using
\ba\label{eq:changeinh}
\delta \ch=\delta p^a \dot q_a - \dot p^a \delta q_a+\delta p^a \delta  \dot q_a \,,
\ea
where the squared term will be important only on 1st order variable redefinitions. In the end, $\delta\ch$ must be a function of $q_a$ and $p^a$ only up to total derivatives. 
A word on notation. Since we take the spatial metric to be flat, i.e. $\delta_{ij}$, we will write only lower indexes for simplicity and we sum over repeated indexes.

Let us expand the Lagrangian up to 3rd order in our variables, which yields
\ba
{\cal L}&=
		\pi_{ij}\dot{\gamma}_{ij}+\pi_E\dot E+\pi_\psi\dot\psi+\pi_\varphi\dot\varphi-\ch_{2}-\ch_{3}\\&
		-A\left(\ch_{N,1}+\ch_{N,2}\right)-\pa_i B\left(\ch_{i,1}+\ch_{i,2}\right)+O(4)
\ea
where we expanded $\ch_N=\sum_{i=1}^3\ch_{N,i}+O(4)$ and $\ch_i=\ch_{i,1}+\ch_{i,2}+O(3)$. Note that $\ch_{2}$ and $\ch_{3}$ differ from $\ch_{N,2}$ and $\ch_{N,3}$ only on total spatial derivatives. See App.\ref{app:perturbhamiltonian} for detailed expressions. Furthermore, we have used that $\ch_{i,0}=0$ and that the background equations of motion hold,\footnote{The Lagrangian starts at 2nd order since the 1st order is proportional to the zeroth order equations of motion} that is
\ba
\ch_{N,0}=-\frac{\pia^2}{12a^3}+\frac{\pip^2}{2a^3}+a^3V=0\,,\\
\dot\pi_\alpha=-6a^3 V \quad {\rm and} \quad \dot\pi_\phi=-a^3V_\phi\,.
\ea
The first and third equations respectively are the first Friedmann equation and the field equations of motion. In the usual notation, they are given by
\ba
3H^2=\frac{1}{2}\dot\phi^2+V\quad{\rm and}\quad \ddot\phi+3H\dot\phi+V_\phi=0\,.
\ea
We decided to keep $\pia$ and $\pip$ since it simplifies the form of the equations, specially when computing $\delta\ch$ from Eq.~\eqref{eq:changeinh}. It is important to note that we have perturbatively expanded in terms of the variables. This does not straightforwardly corresponds to the final perturbation expansion. For example, if we do a 2nd order canonical transformation, e.g. $q_a\to q_a+\delta_2q_a$, what we called $\ch_{2}$ will contain a 3rd order contribution, say $\delta_3\ch_{2}$, which should be added to $\ch_{3}$.

We will proceed using the Faddeev-Jackiw approach \cite{Faddeev:1988qp}. In simple words, we will perturbatively solve the Hamiltonian and Momentum constraints and then we will plug them back into the Lagrangian to find the reduced Hamiltonian. In practice, we need to know the right canonical variables that successfully reduce the system. These variables are gauge invariant, i.e. invariant under coordiante transformation. Thus, before applying the Faddeev-Jackiw method we will first study the gauge transformation of the variables and build the gauge invariant variables up to 2nd order. As it is well known, a particular choice of gauge invariant variables corresponds to a certain slicing. At that point, we can solve the constraints and find the reduced Hamiltonian. We will see this discussion in detail in the following sections.

As we have emphasized in previous sections, the change of variables under coordinate transformation is given by the constraints. In particular, the generalization to non-linear coordinate transformation is given by \cite{Bruni:1996im}
\ba\label{eq:gaugetransformation}
q_a\to q_a+{\rm e}^{\raisebox{-0.05ex}{\textbf{--}}\hspace{-2.2mm}{\cl}_{\epsilon}}q_a= q_a+\pb{q_a}{\epsilon^\mu\ch_{\mu,1+2}}+\frac{1}{2}\pb{\pb{q_a}{\epsilon^\mu\ch_{\mu,2}}}{\epsilon^\nu\ch_{\nu,1}}+O(3)\,,
\ea
where $\epsilon^\mu=\left\{A,\pa_iB\right\}$, $\ch_\mu=\left\{\ch_N,\ch_i\right\}$ and A and B respectively are the time and spatial reparametrization parameters. This is the gauge transformation of our variables up to 2nd order. It is important to note that due to the Poisson algebra the constraints are gauge invariant once the lower order constraints are imposed since we have
\ba
\left\{H_{1+2}[N],H_{1+2}[M]\right\}&=\int d^3x \,a^{-2}\left(N\pa_kM-M\pa_kN\right)\ch_{k,1}+O(3)\,,\\
\left\{H_{i,1+2}[\pa^iB],H_{1+2}[M]\right\}&=\int d^3x \,\pa_kB\pa_kM\,\left(\ch_{N0}+\ch_{N1}\right)+O(3)\,,\\
\left\{H_{i,1+2}[\pa^iB],H_{i,1+2}[\pa^iC]\right\}&=\int d^3x \,\left(\pa_kB\pa_k\pa_lC-\pa_kC\pa_k\pa_lB\right)\,\ch_{l,1}+O(3)\,.
\ea

That is to say that once $\ch_{N1}$ and $\ch_{i,1}$ are solved, $\ch_{N2}$ and $\ch_{i,2}$ are gauge invariant up to 3rd order. Thus, the constraints already provide a way to partially find gauge invariant canonical variables. We will now review the reduction to the 2nd order Hamiltonian and then we will proceed to the 3rd order and find the 2nd order gauge invariant variables.


	\section{2nd order Hamiltonian\label{sec:3}}
	
	Before dealing with the 3rd order Hamiltonian is illustrative to review the 2nd order Hamiltonian and the 1st order gauge invariant variables \cite{Langlois:1994ec}. If we expand the Hamiltonian up to 2nd order in the variables (see App.\ref{app:perturbhamiltonian} for details) we have that, after integration by parts,
	\ba\label{eq:H2}
			\begin{split}
				\ch_{2}&=a^{-3}\left(2\pi_{ij}\pi_{ij}+\frac{3}{4}\left(\Delta^{-1}\pi_E\right)^2-\frac{\pi_{\psi}^2}{12}+\frac{\pi_{\varphi}^2}{2}\right)-\frac{a}{2}\varphi\Delta\vp
				-3\psi\left(H\pi_\psi+\dot{\phi}\,\pi_\varphi\right)\\&
				+a^3 \left(3\psi V_\phi\varphi+\frac{1}{2}V_{\phi\phi}\varphi^2\right)+a\left(\psi-\frac{1}{3}\Delta E\right)\Delta\left(\psi-\frac{1}{3}\Delta E\right)
				-\frac{a}{8}\gamma_{ij}\Delta\gamma_{ij}\,.
	\end{split}
\ea
On the other hand, the Hamiltonian and Momentum constraints are respectively given by
	\ba
		\begin{split}
			\ch_{N,1}=&-\frac{\pia}{6a^3}\pi_\psi +\frac{\pip}{a^3}\pi_\varphi + 6 a^3 \psi V + 
			a^3 \varphi V_\varphi + 2 a \Delta\psi-\frac{2}{3}\Delta\Delta E
		\end{split}
	\ea
	and
	\ba
		\begin{split}
			\pa_i\ch_{i,1}&=\pi_\phi \Delta\varphi+\pi_\alpha \Delta\psi-\frac{1}{3}\Delta\pi_\psi-\pi_E\,.
		\end{split}
	\ea
	Note that since we are interested in going to 3rd order in the Hamiltonian we will not set the constraints to be zero yet. Let us show that indeed the constraints yield the usual gauge transformation rules. The canonical variables transform as
	\ba\label{eq:gaugevariables}
	\psi&\to \psi+\pb{\psi}{\epsilon^b\ch_{b,1}}=\psi-\frac{\pia}{6a^3}A+\frac{1}{3}\Delta B\,,\\
	\vp&\to \vp+\pb{\vp}{\epsilon^b\ch_{b,1}}=\vp+\frac{\pip}{a^3}A\,,\\
	E&\to E+\frac{3}{4}\Delta^{-2}\pa_k\pa_l\pb{Y_{kl}}{\epsilon^b\ch_{b,1}}=E+B\,,\\
	\gamma_{ij}&\to\gamma_{ij}+\widehat{TT}_{ij}\,^{kl}\pb{Y_{kl}}{\epsilon^b\ch_{b,1}}=\gamma_{ij}\,.
	\ea
On the other hand, for the canonical momenta we find
	\ba\label{eq:gaugemomenta}
	\pi_\psi&\to \pi_\psi+\pb{\pi_\psi}{\epsilon^b\ch_{b,1}}=\pi_\psi-6a^3V A-2a\Delta A+\pia\Delta B\,,\\
	\pi_\vp&\to \pi_\vp +\pb{\pi_\vp}{\epsilon^b\ch_{b,1}}=\pi_\vp-a^3V_\phi A+\pip\Delta B\,,\\
	\pi_E&\to \pi_E+2\pa_k\pa_l\pb{P^{kl}}{\epsilon^b\ch_{b,1}}=\pi_E+\frac{2a}{3}\Delta\Delta A\,,\\
	\pi_{ij}&\to \pi_{ij}+\widehat{TT}_{ij}\,^{kl}\pb{P_{kl}}{\epsilon^b\ch_{b,1}}=\pi_{ij}\,.
	\ea
	Detailed formulas on the Poisson brackets can be found in Appendix \ref{app:poissonpert}. We will proceed to construct the gauge invariant variables and then reduce the Hamiltonian. As we previously mentioned, a certain choice of variables corresponds to a particular choice of slicing. In other words, we must choose which variables play the role of the time and spatial slicing parameters $A$ and $B$. For example, we can build several variables which serve as a time slicing, i.e. they only change under a time re-parametrization. The simplest ones are $\vp$, $\pi_E$ and 
		\ba\label{eq:cR}
		\cR\equiv\psi-\frac{1}{3}\Delta E\,.
		\ea
	They respectively correspond to the uniform-$\phi$, shear-free (Newtonian) and flat slicings. The uniform-$\phi$ slicing corresponds to a choice of time coordinate where the scalar field is spatially homogeneous. In a sense, we are setting the perturbations of the inflaton to zero, $\vp=0$, and then all the fluctuations are encoded in the metric. This is useful on super-horizon scales as the curvature perturbation remains constant. The Newtonian slicing is useful at small scales since it reduces to Newtonian gravity in the small scale limit. In this choice of time coordinate, the inflaton fluctuates but there are no fluctuations in the shear, i.e. $\sigma_g=\dot E - B=0$. Thus, these two slicing choices are useful in opposite regimes (super and sub-horizon) and understanding the connection between them is important when super-horizon physics re-enter the horizon.
	
	 In what follows, we will choose $\vp$ and $E$ to represent the temporal and spatial degrees of freedom respectively, i.e. it will be equivalent to work in the uniform-$\phi$ slicing.\footnote{In the literature, the uniform-$\phi$ slicing is sometimes improperly referred to as comoving slicing/gauge, which corresponds to coordinate choice comoving with the total matter. To avoid confusions, we stick to the exact terminology of uniform-$\phi$ slicing.} Later we will come back to the Newtonian slicing.
	
     \subsection{1st order gauge invariance and reduced 2nd order Hamiltonian\label{sec:firstorder}}
		
	Let us find a set of gauge invariant variables that reduce the Hamiltonian up to 2nd order in perturbation. First of all, since the constraints are already gauge invariant, we can use them as canonical variables. Furthermore, any gauge invariant variable commutes with the constraints by definition. Thus, we first choose the gauge invariant variable of our interest and then we reduce the system according to our choice. For simplicity, we chose the gauge invariant variable which coincides with $\psi$ in the uniform-$\phi$ slicing, a.k.a. the comoving curvature perturbation, namely
	\ba\label{eq:1storderzeta}
	\omega=\psi+\frac{\pia}{6\pip}\vp-\frac{1}{3}\Delta E\,.
	\ea
	We will understand this choice as a canonical transformation for the variable $\psi$. Its corresponding canonical momenta is given by
	\ba\label{eq:1stpizeta}
	\pi_\omega=\pi_\psi-\pia \Delta E+\frac{6a^6V}{\pip}\vp+\frac{2a^4}{\pip}\Delta\vp-3\pia\omega-\frac{12a^4}{\pia}\Delta\omega\,,
	\ea
	where the last two terms in the momenta have been introduced to simplify the form of the Hamiltonian. Note that we do not need the terms proportional to $\omega$ for the transformation to be canonical nor for the gauge invariance of $\pi_\omega$. We introduced them for later simplification of the Hamiltonian. Also note that a change $\pi_\omega\to\pi_\omega+f(\omega,t)$, where $f$ is an arbitrary function of $\omega$ and $t$, is equivalent to temporal integration by parts in the action. As it can be seen from the fact that $\pi_\omega\dot\omega\to\pi_\omega\dot\omega+f(\omega)\dot\omega$. The last term can be integrated by parts and could contribute to the Hamiltonian. One can easily see that $\omega$ and $\pi_\omega$ coincide with $\psi$ and $\pi_\psi$ on the uniform-$\phi$ slicing, i.e. $\vp=0$.

	We proceed by defining the constraints as a new canonical momenta for $E$ and $\vp$. Explicitly we have
	\ba\label{eq:pie}
	\pi_{\cal E}\equiv-\pa_i\ch_i=\pi_{E}+\frac{1}{3}\Delta\pi_\psi-\pi_\phi \Delta\varphi-\pi_\alpha\, \Delta\psi
	\ea
	and
	\ba\label{eq:pip}
	\pi_{\delta\phi}\equiv&\frac{a^3}{\pip}\ch_{N1}=\pi_\varphi-\frac{\pia}{6\pip}\pi_\psi  -\frac{6a^6}{\pia}\left(V-\frac{\pia}{6\pip}V_\phi\right)\varphi\\&
	+\frac{6a^6V}{\pip}\left(\psi+\frac{\pip}{\pia}\varphi\right)+\frac{2a^4}{\pip}\Delta\left(\psi-\frac{1}{3}\Delta E\right)\,.
	\ea
	We need not change the variables $E$ and $\vp$ and, thus, we keep the same notation. Tensor modes are already gauge invariant at 1st order. Inverting the canonical transformation we have
	\ba\label{eq:1storder}
	\pi_E&=\pi_{\cal E}-\frac{1}{3}\Delta\pi_\omega-\frac{4a^4}{\pia}\Delta\Delta\omega+\frac{2a^4}{3\pip}\Delta\Delta\vp\\
	\pi_\vp&=\pi_{\dphi}+\frac{\pia}{6\pip}\pi_\omega+3\pip\omega+\pip\Delta {\cal E}-\frac{a^6V_\phi}{\pip}\vp\\
	\pi_\psi&=\pi_\omega+3\pia\omega+\frac{12 a^4}{\pia}\Delta\omega-\frac{6a^6V}{\pip}\vp-\frac{2a^4}{\pip}\Delta\vp+\pia\Delta{\cal E}\,.
\ea
	
	Now that we have our new set of canonical variables, which are gauge invariant at 1st order, we can compute the corresponding Hamiltonian.
First of all, after integration by parts Eq.~\eqref{eq:changeinh}, that is the change in the Hamiltonian, leads us to
	\ba
	\delta\ch&=-\dt{\frac{6 a^6 V}{\pip}}\omega\vp-\dt{\frac{2a^4}{\pip}}\omega\Delta\vp-\dt{\frac{2a^6V}{\pip}}\vp\Delta E
	+\dt{\frac{\pia}{6}}\Delta E \Delta E\\&
	+\dt{\pia}\omega\Delta E+\dt{\frac{3\pia}{2}}\omega^2+ \dt{\frac{6a^4}{\pia}}\omega\Delta\omega-\dt{\frac{a^6}{2\pip}\left(V_\phi-\frac{\pia}{\pip}V\right)}\vp^2\\&
	+\dt{\frac{a^4\pia}{6\pip^2}}\vp\Delta\vp+\dt{\frac{\pia}{6\pip}}\vp\left(\pi_\omega+\pia \Delta E-\frac{6a^6V}{\pip}\vp-\frac{2a^4}{\pip}\Delta\vp+3\pia\omega+\frac{12a^4}{\pia}\Delta\omega\right)\,.
	\ea
	Adding the latter contribution to the original Hamiltonian Eq.~\eqref{eq:H2}, we have that the reduced Hamiltonian at 2nd order is given by
	\ba\label{eq:H2red}
	\ch_2^{\rm red}&=\ch_2+\delta\ch=2a^{-3}\pi_{ij}\pi_{ij}-\frac{a}{8}\gamma_{ij}\Delta\gamma_{ij}+\frac{\pi_\omega^2}{4a^3\epsilon}-a\epsilon\omega\Delta\omega
	+\frac{3}{4a^3}\big(\Delta^{-1}\pi_\mathcal{E}\big)^2+\frac{\pi_{\dph}^2}{2a^3}
	\\&-\pi_{\mathcal{E}}\left(\frac{\Delta^{-1}\pi_\omega}{2a^3}+\frac{6a}{\pia}\omega-\frac{a}{\pip}\vp\right)+\pi_{\dph}\left(\frac{\pia}{6a^3\pip}\pi_\omega+\frac{\pia}{2a^3}\vp-\frac{a^3V_\phi}{\pip}\vp\right)\,,
	\ea
	
	where the super-index red refers to reduced Hamiltonian and we used that
	\ba
	\epsilon\equiv\frac{18\pip^2}{\pia^2}=\frac{1}{2}\frac{\dot\phi^2}{H^2}\,.
	\ea
	Note that if we were to solve the constraints at the moment, they would yield $\pi_{\cal E}=\pi_{\dph}=0$ and $\ch^{\rm red}_2$ would simply be the Hamiltonian for the conserved comoving curvature perturbation, a.k.a. Mukhanov-Sasaki variable \cite{Kodama:1985bj,Mukhanov:1990me}. Nevertheless, since we are interested in the 3rd order Hamiltonian we need to bear in mind that $\pi_{\cal E}$ and $\pi_{\dph}$ contain 2nd order terms in perturbation expansion. Let us apply a similar logic to the 3rd order Hamiltonian.

	\section{3rd order Hamiltonian\label{sec:4}}
	
	After the previous canonical transformation Eqs.~\eqref{eq:1storderzeta} and \eqref{eq:1storder}, the Lagrangian of our system is given by
\ba
	{\cal L}&=
	\pi_{ij}\dot{\gamma}_{ij}+\pi_\omega\dot\omega+\pi_{\cal E}\dot { E}+\pi_{\dphi}\dot\vp-\ch^{\rm red}_{2}-\ch_{3}\\&
	-A\left(\frac{\pip}{a^3}\pi_{\dphi}+\ch_{N,2}\right)-B \left(\pi_{\cal E}-\pa_i\ch_{i,2}\right)+O(4)\,.
\ea

The 3rd order Hamiltonian, after integration by parts, is given by
	\ba
	\ch_{3}&=-6\psi a^{-3}\pi_{ij}\pi_{ij}+9a^{-3}\pi_{ij}\pa_i\psi\pa_j\Delta^{-2}\pi_E-\frac{27}{8a^3}\pa_i\Delta^{-2}\pi_E\pa_j\Delta^{-2}\pi_E\left(\Delta\delta_{ij}-\pa_i\pa_j\right)\psi\\&
	-3a^{-3}\psi\left(\left(\Delta^{-1}\pi_E\right)-\frac{\pi_\psi^2}{12}+\frac{\pi_\vp^2}{2}\right)+\frac{9\psi^2}{2a^3}\left(\pip\pi_\vp-\frac{\pia}{6}\pi_\psi\right)
	+9a^3V\psi^3+\frac{9a^3}{2}V_\phi\vp\psi^2\\&+\frac{3a^3}{2}V_{\phi\phi}\psi\vp^2
	+\frac{a^3}{6}V_{\phi\phi\phi}\vp^3+\frac{a}{2}\cR^2\Delta\cR+\frac{a}{8}\psi\pa_k\gamma_{ij}\pa_k\gamma_{ij}-\frac{1}{4}D_{kl}E \pa_k\gamma_{ij}\pa_l\gamma_{ij}+a\gamma_{ij}\pa_i\cR\pa_j\cR
	\\&-\frac{a}{2}\gamma_{ij}\pa_i\vp\pa_j\vp
	+\frac{a}{2}\cR\pa_k\gamma_{ij}\pa_i\pa_j\pa_k E-\frac{a}{4}\Delta\gamma_{ij}\pa_i\pa_k E\pa_j\pa_k E+2aD_{ij}E\pa_i\cR\pa_j\cR
	-aD_{ij}E\pa_i\vp\pa_j\vp\\&+\frac{a}{2}\psi\pa_i\vp\pa_i\vp+\frac{a}{2}\left(\Delta\delta_{ij}-\pa_i\pa_j\right)\cR\pa_i\pa_kE\pa_j\pa_k E-\frac{a}{8}(\gamma_{kl}\pa_k\gamma_{ij}\pa_l\gamma_{ij}+2\gamma_{ij}\pa_k\gamma_{il}\pa_l\gamma_{jk})\,,
	\ea
	where for simplicity we did not yet apply the transformation Eqs.~\eqref{eq:1storderzeta} and \eqref{eq:1storder} and for convenience we used the shortcut notation for $\cR$ in Eq.~\eqref{eq:cR}. The Hamiltonian and Momentum constraints at 2nd order are respectively given by
	\ba
	\ch_{N,2}&=a^{-3}\left(2\pi_{ij}\pi_{ij}+\frac{3}{4}\left(\Delta^{-1}\pi_E\right)^2-\frac{1}{12}\pi_\psi^2+\frac{1}{2}\pi_\vp^2\right)-3\psi\frac{\pip}{a^3}\left(\pi_\vp-\frac{\pia}{6\pip}\pi_{\psi}\right)
	\\&+a^3\left(3\psi V_\phi\vp+\frac{1}{2}V_{\phi\phi}\varphi^2\right)+a\cR\Delta\cR+a\pa_i\left(\cR\pa_i\cR\right)+\frac{1}{2}\pa_i\vp\pa_i\vp
	+\frac{a}{8}\pa_k\gamma_{ij}\pa_k\gamma_{ij}\\&
	+3a^{-3}\pi_{ij}\pa_i\pa_j\Delta^{-2}\pi_E+\frac{9}{8a^3}\left(\Delta\delta_{ij}-\pa_i\pa_j\right)\pa_i\Delta^{-2}\pi_E\pa_j\Delta^{-2}\pi_E
	-2a\pa_j\left(\Delta E\pa_j\cR\right)\\&-4a\left(\Delta\delta_{ij}-\pa_i\pa_j\right)\pa_i\cR\pa_jE+\frac{a}{2}\left(\Delta\delta_{ij}-\pa_i\pa_j\right)\pa_iE\pa_jE
	-2a\gamma_{ij}\pa_i\pa_j\cR+\frac{a}{2}\pa_i\pa_j\left(\pa_k\gamma_{ij}\pa_k E\right)
	\ea
	and
\ba
	\ch_{i,2}&=
	\pi_\varphi \pa_i\varphi+\pi_\psi \pa_i\psi+\pi^{kl}\pa_i \gamma_{kl}+\frac{3}{4}\pa^k\pa^l\Delta^{-2}\pi_E\pa_i \gamma_{kl}+2\pi^{kl}\pa_i \pa_k\pa_lE+\frac{3}{2}D^{kl}\Delta^{-2}\pi_E\pa_i D_{kl}E\,.
\ea
With these constraints, we can study the gauge transformations at 2nd order and build the 2nd order gauge invariant variables. See App.\ref{app:perturbhamiltonian} for a detailed derivation of the expansion of the Hamiltonian.

	\subsection{2nd order gauge invariance}
	
	The calculations in this subsection are rather involved and, thus, we will provide a handful of them in this section. The detailed formulas can be found in Apps. \ref{app:poissonpert} and \ref{app:canonicalvariables}. Let us focus on the transformation of the two canonical variables $\omega$ and $\gamma_{ij}$. First we will present the results obtained directly from the 2nd order transformation and later we will plug in the solution of the 1st order constraints and remove any redundant gauge invariance. In other words, we want our variables to coincide with $\psi$ and $\gamma_{ij}$ in the uniform-$\phi$ slicing where $\vp=0$. For the canonical scalar variable we have that up to 2nd order transforms according to
	\ba
	\omega\to \omega+\pb{\omega}{\epsilon^\mu\ch_{\mu,1+2}}+\frac{1}{2}\pb{\pb{\omega}{\epsilon^\mu\ch_{\mu,2}}}{\epsilon^\nu\ch_{\nu,1}}\,,
	\ea
	which explicitly yields
	
	\ba
	\omega\to\omega&-\frac{\pi_\psi}{6a^3}A+\frac{\pia}{6\pip}\frac{\pi_\vp}{a^3}A-\frac{\Delta^{-1}}{a^3}\pa_i\pa_j\left(AP^{ij}\right)+\pa_iB\pa_i\psi+\frac{\pia}{6\pip}\pa_iB\pa_i\vp-\frac{1}{4}\Delta^{-1}\pa_i\pa_j\left(\pa_k B\pa_k Y_{ij}\right)\\&
	+\frac{1}{2}A^2\left(V-\frac{\pia}{6\pip}V_\phi\right)+\frac{1}{4a^2}\left(\delta_{ij}-\pa_i\pa_j\Delta^{-1}\right)\left(A\pa_i\pa_jA\right)+\frac{1}{4}\left(\delta_{ij}-\pa_i\pa_j\Delta^{-1}\right)\left(\pa_kB\pa_i\pa_j\pa_kB\right)\,.
	\ea
	After some algebraic manipulations, we find that the scalar gauge invariant variable at 2nd order is given by
	\ba\label{eq:zeta}
	\omega^{GI}&\equiv\omega-\frac{a^6}{2\pi_\phi^2}\left(V-\frac{\pia}{6\pip}V_\phi\right)\varphi^2+\frac{1}{6\pi_\phi}\pi_1\varphi-\pa_i\omega\pa_iE+\frac{\Delta^{-1}}{4}\left(\pa_k\gamma_{ij}\pa_i\pa_j\pa_kE\right)+\frac{\Delta^{-1}}{\pip}\left(\pi_{ij}\pa_i\pa_j\varphi\right)\\&
	+\frac{9}{16a^4}\left(\delta_{ij}-\pa_i\pa_j\Delta^{-1}\right)\left(\pa_i\Delta^{-2}\pi_E\pa_j\Delta^{-2}\pi_E\right)
	+\frac{1}{4}\left(\delta_{ij}-\pa_i\pa_j\Delta^{-1}\right)\left(\pa_i\pa_kE\pa_j\pa_kE\right)\,,
	\ea
	where we defined for simplicity
		\ba
		\pi_1&\equiv\pi_\psi-\frac{\pia}{\pip}\pi_\varphi+3\Delta^{-1}\pi_E+\left(V-\frac{\pia}{6\pip}V_\phi\right)\frac{6a^6}{\pip}\vp
		=3\Delta^{-1}\pi_{\cal E}-\frac{\pia}{\pip}\pi_{\dphi}-\frac{\pia^2}{6\pip^2}\pi_\omega\,,
		\ea
	that is a 1st order gauge invariant variable and the equality holds once the constraints at 1st order are solved.
	On the other hand, the transformation of the tensor modes is given by
	\ba
	\gamma_{ij}\to \gamma_{ij}+\pb{\gamma_{ij}}{\epsilon^\mu\ch_{\mu,1+2}}+\frac{1}{2}\pb{\pb{\gamma_{ij}}{\epsilon^\mu\ch_{\mu,2}}}{\epsilon^\nu\ch_{\nu,1}}\,,
	\ea
	which leads us to
	\ba
	\gamma_{ij}\to\gamma_{ij}+\widehat{TT}_{ij}\,^{ab}\left\{4a^{-3}P_{ab}A+\pa_kB\pa_kY_{ab}+a^{-2}A\pa_a\pa_bA+\pa_kB\pa_a\pa_b\pa_kB\right\}\,.
	\ea
    We find that the gauge invariant tensor modes at 2nd order are given by
	\ba\label{eq:gammaij}
	\gamma_{ij}^{GI}&\equiv\gamma_{ij}+\widehat{TT}_{ij}\,^{ab}\Bigg\{\frac{9}{4a^4}\pa_{(a}\Delta^{-2}\pi_E\pa_{b)}\Delta^{-2}\pi_E-\frac{4}{\pip}\vp\pi_{ab}+\pa_k\pa_{(a}E\pa_{b)}\pa_kE-\pa_k\gamma_{ab}\pa_kE\Bigg\}\,.
	\ea
	Similarly, one can compute the transformation of the canonical momenta and find that the gauge invariant canonical momenta for the scalar and tensor modes are respectively given by
	\ba\label{eq:pizeta}
	\pi_\omega^{GI}&\equiv\pi_\omega+\frac{27}{4\pia}\left(\pa_i\pa_j-\Delta\delta_{ij}\right)\pa_i\Delta^{-2}\pi_E\pa_j\Delta^{-2}\pi_E-\frac{12a^4}{\pia\pip}\pi_{ij}\pa_i\pa_j\vp-\frac{3a^4}{\pia}\pa_k\gamma_{ij}\pa_i\pa_j\pa_kE+...\,,
	\ea
	where the full expression can be found in App.~\ref{app:canonicalvariables}, and
	\ba\label{eq:piij}
	\pi_{ij}^{GI}\equiv\pi_{ij}+\widehat{TT}_{ij}\,^{ab}\Bigg\{&\frac{2a^4}{\pip}\pa_{(a}\omega\pa_{b)}\vp-\frac{a^4\pia}{6\pip^2}\pa_{a}\vp\pa_{b}\vp-\frac{3}{4}\pa_k\left(\pa_a\pa_b\Delta^{-2}\pi_E\pa_kE\right)\\&
	-\pa_k\left(\pi_{ab}\pa_k E\right)-\frac{a^4}{4\pip}\pa_k\left(\vp\pa_k\gamma_{ab}\right)\Bigg\}\,.
	\ea

	So far we have derived some 2nd order gauge invariant variables directly from the coordinate transformations. However, it should be noted that the derived 2nd order gauge invariant variables, Eqs.~\eqref{eq:zeta}, \eqref{eq:gammaij}, \eqref{eq:pizeta} and \eqref{eq:piij} contain a spurious gauge invariance once the 1st order constraints are imposed in the uniform-$\phi$ slicing. For example, take a look at Eq.~\eqref{eq:zeta}. The term with two first derivatives of $\Delta^{-2}\pi_E$ will yield terms proportional to $O(\pi_\omega^2)$ and $O(\omega^2)$ once Eq.~\eqref{eq:1storder} is used. Not to alter the definition of the non-linearly conserved curvature perturbation, we must remove such redundant terms. 	Thus, after imposing Eq.~\eqref{eq:1storder} and removing the redundant gauge invariant terms we find that the canonical variables are given by
	
	\ba\label{eq:zeta2}
	\zeta\equiv\omega&+\frac{\eta}{8\epsilon}\varphi^2-\frac{1}{2\epsilon\pi_\phi}\pi_\omega\varphi-\pa_i\omega\pa_iE+\frac{\Delta^{-1}}{4}\left(\pa_k\gamma_{ij}\pa_i\pa_j\pa_kE\right)+\frac{\Delta^{-1}}{\pip}\left(\pi_{ij}\pa_i\pa_j\varphi\right)\\&
	+\frac{1}{4}\left(\delta_{ij}-\pa_i\pa_j\Delta^{-1}\right)\left(\pa_i\pa_kE\pa_j\pa_kE+\frac{a^4}{\pip^2}\pa_i\vp\pa_j\vp-\frac{12a^4}{\pip\pia}\pa_i\vp\pa_j\omega-\frac{1}{\pip}\pa_i\vp\pa_j\Delta^{-1}\pi_\omega\right)
	\ea
	and
		\ba\label{eq:gamma2}
		\gamma^C_{ij}\equiv\gamma_{ij}+\widehat{TT}_{ij}\,^{ab}&\Bigg\{-\frac{4}{\pip}\vp\pi_{ab}
		+\pa_k\pa_{(a}E\pa_{b)}\pa_kE-\pa_k\gamma_{ab}\pa_kE
		+\frac{a^4}{\pip^2}\pa_a\vp\pa_b\vp\\&-\frac{12a^4}{\pip\pia}\pa_a\vp\pa_b\omega-\frac{1}{\pip}\pa_a\vp\pa_b\Delta^{-1}\pi_\omega\Bigg\}\,,
		\ea
	
	where the subindex $C$ refers to the fact that the variable corresponds to the one in the uniform-$\phi$ slicing and in Eq.~\eqref{eq:zeta2} we used that
	\ba
	\eta\equiv\frac{\dot \epsilon}{H\epsilon}=-72\frac{a^6}{\pia^2}\left(V-\frac{\pia}{6\pip}V_\phi\right)\,.
	\ea
	In passing, note that the term proportional to $\vp^2$ in Eq.~\eqref{eq:zeta2} exactly matches that from the $\delta N$ formalism on superhorizon scales \cite{Sasaki:1995aw}. To compare with the literature, we can set $E=0$ and treat Eqs.~\eqref{eq:zeta2} and \eqref{eq:gamma2} as the relation between the uniform-$\phi$ and flat slicings. For example, one can easily check that this formula coincides with Ref.~\cite{Maldacena:2002vr}. Although there is a difference in the last term of the first line of Eq.~\eqref{eq:zeta2} (the last term in Eq.~(A.8) in Ref.~\cite{Maldacena:2002vr}), namely the non-local operator is missing. Our expression also matches that of Malik \& Wands \cite{Malik:2008im} after a proper redefinition of the variable, concretely $\zeta=\zeta^{MK}_1+\frac{1}{2}\zeta^{MK}_2-\left(\zeta^{MK}_1\right)^2$ bearing in mind that we work in proper cosmic time and that our definition of tensor modes carries an extra factor $2$.

	The definitions of the corresponding canonical momenta for $\zeta$ is given by
	\ba\label{eq:pizeta2}
	\pi_\zeta\equiv\pi_\omega&+\frac{3a^4}{\pia\pip}\left(\Delta\delta_{ij}-\pa_i\pa_j\right)\left[\pa_i\vp\pa_j\Delta^{-1}\pi_\omega+\frac{12a^4}{\pia}\pa_i\vp\pa_j\omega-\frac{a^4}{\pip}\pa_i\vp\pa_j\vp\right]\\&-\frac{12a^4}{\pia\pip}\pi_{ij}\pa_i\pa_j\vp
	+\frac{9}{2}\pia\zeta^2+...\,,	
	\ea
where the full expression is presented in App.~\ref{app:canonicalvariables}. Note that we added an extra $\zeta^2$ term with the same purpose as in Eq.\eqref{eq:1stpizeta} to simplify the form of the Hamiltonian. The canonical momenta for the tensor modes reads
	\ba\label{eq:piij2}
	\pi^C_{ij}\equiv\pi_{ij}+\widehat{TT}_{ij}\,^{ab}\Bigg\{&\frac{2a^4}{\pip}\pa_{(a}\omega\pa_{b)}\vp-\frac{a^4\pia}{6\pip^2}\pa_{a}\vp\pa_{b}\vp+\frac{1}{4}\pa_k\left(\pa_a\pa_b\Delta^{-1}\pi_\omega\pa_kE\right)+\frac{3a^4}{\pia}\pa_k\left(\pa_a\pa_b\omega\pa_kE\right)\\&
	-\frac{a^4}{2\pip}\pa_k\left(\pa_a\pa_b\vp\pa_kE\right)-\pa_k\left(\pi_{ab}\pa_k E\right)-\frac{a^4}{4\pip}\pa_k\left(\vp\pa_k\gamma_{ab}\right)\Bigg\}\,.
	\ea
	\newpage

	This set of canonical variables, Eqs.~\eqref{eq:zeta2}-\eqref{eq:piij2}, are the ones we use for the reduction of the Hamiltonian and that coincide with the variables in the uniform-$\phi$ slicing. Before ending this section, it is important to note that since $E$ and $\vp$ only appear as 3rd order combinations or are always multiplied by $\pi_{\cal E}$ and $\pi_{\dph}$, we do not need to redefine them even at 3rd order. Just as we did at 1st order in perturbation, we will redefine one canonical scalar variable, in our case $\omega$, and the corresponding momenta $\pi_\omega$. 
	
\subsection{3rd order reduced Hamiltonian}
	We are left to perform the canonical transformations Eqs.~\eqref{eq:zeta2}-\eqref{eq:piij2} and reduce the Hamiltonian. Note that we did not use the constraints to define new momenta. Actually, if we were interested in the 4th order Hamiltonian we would redefine a new momenta for $E$ and $\vp$. However, since we only study the system up to 3rd order we will solve the Hamiltonian and Momentum constraints respectively leading us to
	\ba\label{eq:2ndorderconstraints}
	\pi_{\dphi}=-\frac{a^3}{\pip}\ch_{N,2}\quad {\rm and}\quad\pi_{\cal E}=\pa_i\ch_{i,2}\,.
	\ea
In doing so, we are solving the constraints perturbatively, e.g. we replace the 1st order solutions Eq.~\eqref{eq:1storder} in $\ch_{N,2}$ and $\ch_{i,2}$. One can check that the error made is higher order. 

First, we compute the change in the Hamiltonian from the time dependent canonical transformation. Since the algebra is heavy, specially for the scalar modes, let us show here the result for terms containing two tensor modes. The complete results can be found in App.\ref{app:changelagrangian}. The change in the Hamiltonian is given by
	\ba\label{eq:H33}
	\delta\ch&=-\dt{\frac{2}{\pip}}\pi_{ij}^C\pi_{ij}^C\vp-\dt{\frac{a^4}{8\pip}}\vp\pa_k{\gamma_{ij}^C}\pa_k{\gamma_{ij}^C}+...
	\ea
	On the other hand, from the reduced 2nd order Hamiltonian \eqref{eq:H2red} we have a 3rd order contribution after the transformation as well. In the present case, it reads
	 
		\ba
		\delta_3\ch_2^{\rm red}=&-4a^{-3}\pi^C_{ij}\delta\pi_{ij}+\frac{a}{4}\delta\gamma_{ij}\Delta\gamma^C_{ij}-\frac{\pi_\zeta}{2a^3\epsilon}\delta\pi_{\omega}+2a\epsilon\zeta\Delta\delta\omega
		-\pa_i\ch_{i,2}\left(\frac{\Delta^{-1}\pi_\zeta}{2a^3}+\frac{6a}{\pia}\zeta-\frac{a}{\pip}\vp\right)
		\\&-\frac{a^3}{\pip}\ch_{N,2}\left(\frac{\pia}{6a^3\pip}\pi_\zeta+\frac{\pia}{2a^3}\vp-\frac{a^3V_\phi}{\pip}\vp\right)\,,
		\ea
where $\delta q_a=q_a^{GI}-q_a$ and we have used Eq.~\eqref{eq:2ndorderconstraints}. Again, extracting only the terms containing two tensor modes we find
\ba\label{eq:H32}
	\delta_3\ch_2^{\rm red}&=-\frac{\pia}{3a^3\pip^2}\pi_\zeta\pi^C_{ij}\pi^C_{ij}+\frac{2}{a^3\pip}\left(\frac{a^6V_\phi}{\pip}-\frac{\pia}{2}\right)\pi^C_{ij}\pi^C_{ij}\vp+2a^{-3}\pi^C_{ij}\pi^C_{ij}\Delta E\\&+\frac{a}{8\pip}\left(\frac{a^6V_\phi}{\pip}-\frac{\pia}{2}\right)\vp\pa_k\gamma^C_{ij}\pa_k\gamma^C_{ij}
	-\frac{a\pia}{64\pip^2}\pi_\zeta\pa_k\gamma^C_{ij}\pa_k\gamma^C_{ij}-\frac{a}{4}\Delta\gamma^C_{ij}\pa_k\gamma^C_{ij}\pa_kE
	\\&+\frac{6a}{\pia}\pi^C_{ij}\pa_{k}\gamma^C_{ij}\pa_k\zeta+\frac{1}{2a^3}\pi_{ij}^C\pa_{k}\gamma^C_{ij}\pa_k\Delta^{-1}\pi_\zeta+...
\ea
For the 3rd order Hamiltonian piece we find that the terms containing two tensor modes are given by
\ba\label{eq:H31}
\ch_3=&-6a^{-3}\psi\pi^C_{ij}\pi^C_{ij}+\frac{a}{8}\psi\pa_k\gamma^C_{ij}\pa_k\gamma^C_{ij}-\frac{a}{4}\pa_k\pa_l E\pa_k\gamma^C_{ij}\pa_l\gamma^C_{ij}+\frac{a}{12}\Delta E\pa_k\gamma^C_{ij}\pa_k\gamma^C_{ij}+...
\ea
Adding all the 3rd order terms that contain two tensor modes, Eqs.~\eqref{eq:H33}, \eqref{eq:H32} and \eqref{eq:H31}, we find that the reduced 3rd order Hamiltonian is given by
	\ba\label{eq:H3red}
	\ch_3^{\rm red}=&\ch_{3}+\delta_3\ch_2^{\rm red}+\delta\ch=
	-6a^{-3}\zeta\pi^C_{ij}\pi^C_{ij}-\frac{\pia}{3a^3\pip^2}\pi_\zeta\pi^C_{ij}\pi^C_{ij}-\frac{a\pia}{64\pip^2}\pi_\zeta\pa_k\gamma^C_{ij}\pa_k\gamma^C_{ij}\\&+\frac{a}{8}\zeta\pa_k\gamma^C_{ij}\pa_k\gamma^C_{ij}+\pi_{ij}^C\pa_k\gamma^C_{ij}\pa_k\chi+...
	\ea

		where for simplicity we have introduced
		\ba\label{eq:chi}
		\chi\equiv\frac{3}{2a^3}\left(\frac{1}{3}\Delta^{-1}\pi_\zeta+\frac{4a^4}{\pia}\zeta\right)\,.
		\ea
 Note how the terms containing $E$ and $\vp$ exactly cancel. Also note that $\chi$ is proportional to the shift vector in the uniform-$\phi$ slicing, see Ref.~\cite{Maldacena:2002vr}. Regarding the reduction containing all terms, one can check that after a series of manipulations the terms proportional to $E$, $\vp$ completely vanish and the reduced complete 3rd order action is given by
 \ba\label{eq:CL}
 {\cal L}&=
 \pi^C_{ij}\dot{\gamma}^C_{ij}+\pi_\zeta\dot\zeta-\ch^{\rm red}_{2}-\ch^{\rm red}_{3}+O(4)\,,
 \ea
  
 where 
 	\ba\label{eq:H2red2}
 	\ch_2^{\rm red}&=2a^{-3}\pi^C_{ij}\pi^C_{ij}-\frac{a}{8}\gamma^C_{ij}\Delta\gamma^C_{ij}+\frac{\pi_\zeta^2}{4a^3\epsilon}-a\epsilon\zeta\Delta\zeta\,,
 	\ea
 and the reduced 3rd order Hamiltonian reads
	\ba
	\ch_3^{\rm red}&=
	\frac{1}{8a^6H\epsilon^2}\pi_\zeta^3-\frac{3}{4a^3\epsilon}\pi_\zeta^2\zeta+\frac{a}{2}\zeta^2\Delta\zeta
	-\frac{1}{2\epsilon Ha^3}\pi_\zeta\pa_i\zeta\pa_i\zeta
	+\frac{1}{2\epsilon Ha^3}\pi_\zeta\Delta\zeta^2+2a^3\Delta\chi\pa_i\zeta\pa_i\chi\\&
	-\frac{3a^3}{2}\left(\zeta+\frac{\pi_\zeta}{18a^6}\right)\left(\Delta\delta_{ij}-\pa_i\pa_j\right)\pa_i\chi\pa_j\chi-6\pi_{ij}^C\pa_i\zeta\pa_j\chi-6a^{-3}\zeta\pi_{ij}^C\pi_{ij}^C+a\gamma^C_{ij}\pa_i\zeta\pa_j\zeta
	\\&+\frac{1}{\epsilon Ha^2}\left(\frac{1}{a}\pi_{ij}^C\pa_i\chi\pa_j\pi_\zeta+\gamma^C_{ij}\pa_i\zeta\pa_j\pi_\zeta+\frac{1}{a^4}\pi_\zeta\pi_{ij}^C\pi_{ij}^C+\frac{1}{16}\pi_\zeta\pa_k\gamma^C_{ij}\pa_k\gamma^C_{ij}\right)-\frac{a^3}{4}\Delta\gamma^C_{ij}\pa_i\chi\pa_j\chi\\&
	+\frac{a}{8}\zeta\pa_k\gamma^C_{ij}\pa_k\gamma^C_{ij}+\pi_{ij}^C\pa_k\gamma^C_{ij}\pa_k\chi
	-\frac{a}{8}\left(\gamma^C_{kl}\pa_k\gamma^C_{ij}\pa_l\gamma^C_{ij}+2\gamma^C_{ij}\pa_k\gamma^C_{il}\pa_l\gamma^C_{jk}\right)\,.
	\ea

	This Hamiltonian coincides with the plain expansion of the 3rd order action in the uniform-$\phi$ slicing as expected. It can easily be seen if one uses the 1st order identification 
	\ba
	\pi_\zeta=2a^3\epsilon\dot\zeta\quad{\rm and}\quad\pi_{ij}^C=\frac{1}{4}a^3\dot\gamma^C_{ij}
	\ea
	and the fact that $H_3=-L_3$. For example see Eqs.~(3.7) and (3.14) of Ref.~\cite{Maldacena:2002vr}. Thus, we have performed a successful reduction of the system within the Hamiltonian formalism that coincides with that in the uniform-$\phi$ slicing. Before ending this section, let us note that we could proceed as in Ref.~\cite{Maldacena:2002vr} and show that the 3rd order action is of $O(\epsilon^2)$. To show this, we could either work at the action level and integrate by parts or simply do a series of canonical transformations.
	
	\section{Newtonian slicing \label{sec:newtonian}}
	On super-horizon scales, the most used slicing is the uniform-$\phi$ slicing. The reason being that, the comoving curvature perturbation is conserved on super-horizon scales. On sub-horizon scales, one usually focuses on the flat or Newtonian slicing to gain intuition. For example, GWs sourced by 2nd order scalar perturbations are usually computed in the Newtonian slicing.
	For completeness, we show in this section how to move from the uniform-$\phi$ to the Newtonian slicing within the Hamiltonian formalism and in passing we show how the scalar and tensor mix in such a gauge transformation. The Newtonian slicing is a shear free slicing and, thus, we have to use $\pi_E$ as our time reparametrization variable. At the end of this section we recover the well-known equations of motion for tensor modes with 2nd order scalar source terms.
	
	\subsection{2nd order Hamiltonian}
	Proceeding similarly as in Sec.~\ref{sec:firstorder}, we build a gauge invariant variable that coincides with $\psi$ on the shear-free slices. Such variable will correspond to the usual Newtonian potential. For the moment, let us define it as
	\ba
	\theta\equiv\psi-\frac{1}{3}\Delta E+\frac{\pia}{4a^4}\Delta^{-2}\pi_E\,,
	\ea
where note that we used $\pi_E$ as our time slicing parameter. Using Eqs.~\eqref{eq:zeta} and \eqref{eq:xi} we can relate $\theta$ and $\omega$ at linear order by
	\ba
	\theta=\omega-\frac{\pia}{6\pip}\Xi\,,
	\ea
where
	\ba\label{eq:xi}
	\Xi\equiv\vp-\frac{3\pip}{2a^4}\Delta^{-2}\pi_E=\frac{\pip}{2a^4}\Delta^{-1}\pi_\omega+6\frac{\pip}{\pia}\omega\,.
	\ea
Note that in the last step we used the 1st order constraint equations Eq.~\eqref{eq:1storder}.
Thus, the linear transformation between $\omega$ and $\theta$ and its conjugate momenta\footnote{We chose the definition of $\pi_\theta$ so that the resulting Hamiltonian is of the simplest form and coincides with that of Ref.~\cite{Makino:1991sg}.} is respectively given by
	\ba
	\theta=-\frac{\pia}{12a^4}\Delta^{-1}\pi_\omega 
\quad
{\rm and}
\quad
	\pi_\theta=\left(1+\frac{\pia^2}{18\pip^2}\right)\pi_\omega+\frac{12a^4}{\pia}\Delta\omega\,.
	\ea
	After inverting we are led to
	\ba\label{eq:omegatheta}
	\omega=\left(1+\frac{\pia^2}{18\pip^2}\right)\theta+\frac{\pia}{12a^4}\Delta^{-1}\pi_\theta
\quad
{\rm and}
\quad
	 \pi_\omega=-\frac{12a^4}{\pia}\Delta\theta\,.
	\ea
 With this canonical transformation the reduced Hamiltonian is given by
	\ba\label{eq:hamnewton}
	\ch_2^{\rm red, N}=-\frac{\pip^2}{8a^7}\pi_\theta\Delta^{-1}\pi_\theta+a\frac{\eta}{\epsilon}\theta\Delta\theta+\frac{2a^5}{\pip^2}\Delta\theta\Delta\theta
	\ea
	which coincides with that of Ref.~\cite{Makino:1991sg} except for a difference in the coefficient $\theta\Delta\theta$. However, our Hamiltonian Eq.~\eqref{eq:hamnewton} yields the correct equations of motion which can be found in Eq.~(6.48) in Ref.~\cite{Mukhanov:1990me}. Also note that Eq.~\eqref{eq:omegatheta} is the well known linear relation between the comoving curvature perturbation and the Newtonian potential \cite{Mukhanov:1990me}, once the Hamiltonian equations of motion are used. For later convenience we have that $\Xi$ in the Newtonian slicing is given by\footnote{Note that $\Xi$ in Eq.~\eqref{eq:xi} is proportional to the scalar part of the shift vector in the uniform-$\phi$ slicing, namely $\chi$ in Eq.~\eqref{eq:chi}.}
	\ba
	\Xi=\frac{\pia}{3\pip}\theta+\frac{\pip}{2a^4}\Delta^{-1}\pi_\theta=-\frac{2}{\dot\phi}\left(\dot\theta+H\theta\right)\,,
	\ea
	where in the last equality we have used the Hamilton equations of motion.
	\subsection{3rd order Hamiltonian for tensor modes}
	Not to overload the paper with formulas, we will only focus on the mixing terms with the tensor modes. Similar procedure can be applied to the pure scalar sector. Let us start with the definition of tensor modes in the Newtonian (shear free) slicing. They are given by
	
		\ba\label{eq:gammaij3}
		\gamma_{ij}^{N}&\equiv\gamma_{ij}+\widehat{TT}_{ij}\,^{ab}\Bigg\{\frac{9}{4a^4}\pa_{(a}\Delta^{-2}\pi_E\pa_{b)}\Delta^{-2}\pi_E-\frac{6}{a^4}\Delta^{-2}\pi_E\,\pi_{ab}+\pa_k\pa_{(a}E\pa_{b)}\pa_kE-\pa_k\gamma_{ab}\pa_a\pa_b\pa_kE\Bigg\}\,,
		\ea
	where the super-index $N$ refers to Newtonian.
	Similarly, the conjugate momenta is
		\ba\label{eq:piij3}
		\pi_{ij}^{N}\equiv\pi_{ij}+\widehat{TT}_{ij}\,^{ab}\Bigg\{3\pa_{(a}\theta\pa_{b)}\Delta^{-2}\pi_E&-\frac{3\pia}{8a^4}\pa_{a}\Delta^{-2}\pi_E\pa_{b}\Delta^{-2}\pi_E-\frac{3}{4}\pa_k\left(\pa_a\pa_b\Delta^{-2}\pi_E\pa_kE\right)\\&
		-\pa_k\left(\pi_{ab}\pa_k E\right)-\frac{3}{8}\pa_k\left(\Delta^{-2}\pi_E\pa_k\gamma_{ab}\right)\Bigg\}\,.
		\ea
On the other hand, we find that the gauge invariant Newtonian potential at 2nd order is given by
\ba\label{eq:Phi2}
\Phi=\theta+&\frac{\Delta^{-1}}{4}\left(\pa_k\gamma_{ij}\pa_i\pa_j\pa_kE\right)+\frac{3}{2a^4}\Delta^{-1}\left(\pi_{ij}\pa_i\pa_j\Delta^{-2}\pi_E\right)+\frac{\pia}{4a^4}\Delta^{-1}\left(\gamma_{ij}\pa_i\pa_j\Delta^{-2}\pi_E\right)\\&
-\frac{3\pia}{16a^4}\Delta^{-2}\pa_k\left(\pa_k\gamma_{ij}\pa_i\pa_j\Delta^{-2}\pi_E\right)-\frac{\pia}{2a^4}\Delta^{-2}\pa_k\left(\pi_{ij}\pa_i\pa_j\pa_k E\right)+O(\theta^2)\,,
\ea
where we have not taken into account scalar squared contributions, i.e. $O(\theta^2)$. For the tensor modes, we can easily relate the definitions between the uniform-$\phi$ and Newtonian slicing as
			\ba\label{eq:gammaCN}
			\gamma_{ij}^{C}=\gamma_{ij}^{N}+\widehat{TT}_{ij}\,^{ab}\left\{-\frac{4}{\pip}\Xi\pi^N_{ab}-\frac{a^4}{\pip^2}\pa_a\Xi\pa_b\Xi\right\}
			\ea
			and
			\ba\label{eq:piCN}
			\pi_{ij}^{C}=\pi_{ij}^{N}+\widehat{TT}_{ij}\,^{ab}\left\{-\frac{a^4}{4\pip}\pa_k\left(\Xi\pa_k\gamma^N_{ab}\right)+\frac{2a^4}{\pip}\pa_{(a}\Phi\pa_{b)}\Xi+\frac{a^4\pia}{6\pip^2}\pa_a\Xi\pa_b\Xi\right\}\,.
			\ea
			
We can do similar steps for the scalar modes. One possibility is to find the conjugate momenta of $\theta$ in terms of the original variables at linear order, find the gauge invariant momenta at 2nd order and then compute the difference with $\zeta$ and $\pi_\zeta$. Instead, we will use the fact that we know the relation between $\gamma_{ij}$ and $\pi_{ij}$, i.e. Eqs.~\eqref{eq:gammaCN} and \eqref{eq:piCN}, and we will require that the transformation of the scalar variables is canonical.\footnote{In other words, we want $\delta \ch$ to be just a function of the phase space variables.} For our purposes, this approach sufficient. The corresponding relation for the scalar variables is respectively given by

	\ba\label{eq:ztop}
	\zeta=\left(1+\frac{\pia^2}{18\pip^2}\right)\Phi+\frac{\pia}{12a^4}\Delta^{-1}\pi_\Phi&+\frac{\Delta^{-1}}{\pip}\left(\pi^N_{ij}\pa_i\pa_j\Xi\right)+\Delta^{-1}\left(\gamma^N_{ij}\pa_i\pa_j\Phi\right)\\&
	-\frac{\Delta^{-1}}{a^4}\left(\pi^N_{ij}\pi^N_{ij}\right)-\frac{\Delta^{-1}}{16}\left(\pa_k\gamma^N_{ij}\pa_k\gamma^N_{ij}\right)+O(\Phi^2)
	\ea
	and
	\ba\label{eq:pztop}
	\pi_\zeta=-\frac{12a^4}{\pia}\Delta\Phi-&\frac{2a^4}{\pip}\gamma^N_{ij}\pa_i\pa_j\Xi-\frac{12a^4}{\pip\pia}\pi^N_{ij}\pa_i\pa_j\Xi-\frac{12a^4}{\pia}\gamma^N_{ij}\pa_i\pa_j\Phi\\&+\frac{12}{\pia}\pi^N_{ij}\pi^N_{ij}+\frac{3a^4}{4\pia}\pa_k\gamma^N_{ij}\pa_k\gamma^N_{ij}+O(\Phi^2)\,,
	\ea

where again for simplicity we have neglected scalar squared contributions, that is $O(\Phi^2)$. After plugging Eqs.~\eqref{eq:gammaCN}--\eqref{eq:pztop} into the Lagrangian Eq.~\eqref{eq:CL}, see App.\ref{app:changelagrangian2} for the change in the Hamiltonian, we obtain that the reduced Hamiltonian in the Newtonian slicing for the mixing terms up to 3rd order is given by
	 \ba
	 {\cal L}&=
	 \pi^N_{ij}\dot{\gamma}^N_{ij}+\pi_\Phi\dot\Phi-\ch^{\rm red,N}_{2}-\ch^{\rm red,N}_{3}+O(4)\,,
	 \ea
	 where
		\ba
		\ch_2^{\rm red, N}&=2a^{-3}\pi^N_{ij}\pi^N_{ij}-\frac{a}{8}\gamma^N_{ij}\Delta\gamma^N_{ij}-\frac{\pip^2}{8a^7}\pi_\Phi\Delta^{-1}\pi_\Phi+a\frac{\eta}{\epsilon}\Phi\Delta\Phi+\frac{2a^5}{\pip^2}\Delta\Phi\Delta\Phi
		\ea
		and

	\ba
	\ch_3^{\rm red, N}&=-8a^{-3}\Phi\,\pi^N_{ij}\pi^N_{ij}-a\gamma^N_{ij}\pa_i\Phi\pa_j\Phi-\frac{a}{2}\gamma^N_{ij}\pa_i\Xi\pa_j\Xi
	+O(\Phi^3)\,.
	\ea
	As a final check, it is easy to see that the equations of motion for the tensor modes with a scalar source, i.e.
	\ba\label{eq:gwnewton}
	\ddot\gamma^N_{ij}+&3H\dot\gamma^N_{ij}-a^{-2}\Delta\gamma^N_{ij}=\widehat{TT}_{ij}\,^{ab}\Bigg\{4a^{-2}\pa_a\Phi\pa_b\Phi+\frac{8}{a^2\dot\phi^2}\pa_a\left(\dot\Phi+H\Phi\right)\pa_b\left(\dot\Phi+H\Phi\right)\Bigg\}\,,
	\ea
	coincides with Eq.~(A.8) of Ref.~\cite{Alabidi:2012ex} (see also Refs.~\cite{Baumann:2007zm,Ananda:2006af}). We have thus successfully moved from the uniform-$\phi$ to the Newtonian slicing with the proper definition of tensor and scalar modes. Note that one could also check that the same equations of motion are obtained by computing the e.o.m. in the uniform-$\phi$ slicing and plugging in the relation Eq.~\eqref{eq:gammaCN}. We would like to mention that the equations of motion and the 3rd action looks remarkably simpler in the Newtonian slicing than the uniform-$\phi$ slicing.

It is worth estimating the size of the induced tensor modes from 2nd order scalar sources. From Eq.~\eqref{eq:gwnewton} we roughly have that $\gamma^{\rm ind}_k\sim\Phi^2$.  Recall that from quantum fluctuations during inflation we have $\gamma^{\rm inf}_k\sim H/M_{pl}$. Thus, at large scales the 1st order contribution dominates as long as $\epsilon>P_\Phi$, since $\Phi\sim H/(\sqrt{\epsilon}M_{pl})$. It is when the scalar modes re-enter the horizon during radiation domination that the induced tensor modes might overcome those produced during inflation. Specially interesting is the case of Primordial Black Holes, where the power spectrum at small scales might be as large as $P_\Phi\sim 10^{-3}\,.$\,
For example, for $M_{PBH}\sim10^{22}g$ we have $k\sim 10^{12} {\rm Mpc}^{-1}$ and a frequency of $\nu \sim 10^{-3} \rm Hz$ \cite{Alabidi:2012ex}.
 In this way, the GWs energy density today can be estimated to be $\Omega^{\rm ind}_{GW}\propto P_\Phi^2(k)/({1+z_{eq}})$, where $z_{eq}\approx3300$ is the red-shift at matter-radiation equality  \cite{Baumann:2007zm,Alabidi:2012ex}. Plugging in our estimates we are led to $\Omega^{\rm ind}_{GW}\sim10^{-10}$ which falls within the range of space-based GWs detectors (see Ref.~\cite{Romano:2016dpx} and references therein). It should be noted that any 3rd or higher order contribution will be suppressed by an extra factor $P_\Phi^{1/2}$ and thus it would be irrelevant.

	\section{Conclusions\label{sec:5}}
	
	We have devoted this work to clarify the definition of scalar and tensor modes up to 2nd order in perturbation theory in the Hamiltonian formalism. Like in gauge theory, the constraints are the generators of the symmetries (in our case diffeomorphisms) and, thus, the Hamiltonian is suitable to study the gauge issues of cosmological perturbations. We obtained the 2nd order gauge invariant definition of scalar and tensor modes that coincide with scalar and tensor perturbations in the uniform-$\phi$ slicing. With such definitions we reduced the system and obtained for the first time the reduced Hamiltonian at 3rd order in complete generality without any gauge fixing. To summarize, the canonical variables that are gauge invariant at 2nd order and coincide with the curvature perturbation and tensor perturbations in the uniform-$\phi$ slicing are given by
	
	\ba
	\zeta\equiv&\omega+\frac{\eta}{8\epsilon}\varphi^2-\frac{1}{\dot\phi}\vp\dot\omega-\pa_i\omega\pa_iE+\frac{\Delta^{-1}}{4}\left(\pa_k\gamma_{ij}\pa_i\pa_j\pa_kE\right)+\frac{\Delta^{-1}}{4\dot\phi}\left(\dot\gamma^{ij}\pa_i\pa_j\varphi\right)\\&
	+\frac{1}{4}\left(\delta_{ij}-\pa_i\pa_j\Delta^{-1}\right)\left(\pa_i\pa_kE\pa_j\pa_kE+\frac{1}{a^2\dot\phi^2}\pa_i\vp\pa_j\vp+\frac{2}{a^2\dot\phi H}\pa_i\vp\pa_j\omega-\frac{2\epsilon}{\dot\phi}\pa_i\vp\pa_j\Delta^{-1}\dot\omega\right)
	\ea
and
	\ba
	\gamma_{ij}^{C}\equiv\gamma_{ij}+\widehat{TT}_{ij}\,^{ab}\Bigg\{-\frac{1}{\dot\phi}\vp\dot\gamma_{ab}
	&+\pa_k\pa_{(a}E\pa_{b)}\pa_kE-\pa_k\gamma_{ab}\pa_kE
	+\frac{1}{a^2\dot\phi^2}\pa_a\vp\pa_b\vp\\&+\frac{2}{a^2\dot\phi H}\pa_a\vp\pa_b\omega-\frac{2\epsilon}{\dot\phi}\pa_a\vp\pa_b\Delta^{-1}\dot\omega\Bigg\}\,,
	\ea
	where
$
	\omega=\psi-\frac{H}{\dot\phi}\vp-\frac{1}{3}\Delta E\,.
$
	Note that $\omega$ coincides with the comoving curvature perturbation at the linear level.

For completeness we checked that the relation between the uniform-$\phi$ and flat slicing coincide with the already existing results in the literature, e.g. Refs.~\cite{Maldacena:2002vr,Malik:2008im}. After the reduction of the system, the reduced Hamiltonian coincides with the straightforward expansion of the action and, thus, one can apply the known simplification to show that the action is of order $O(\epsilon^2)$, e.g. Ref.~\cite{Maldacena:2002vr}. Alternatively, in the Hamiltonian formalism one can perform 2nd order canonical transformations until the 3rd order Hamiltonian is $O(\epsilon^2)$ in the uniform-$\phi$ slicing.
	
	To extend the discussion, we showed how one can move from the uniform-$\phi$ to the Newtonian slicing. For simplicity, we considered only the mixing terms between scalar and tensor modes. We find that the 2nd order gauge invariant comoving curvature perturbation $\zeta$ is related to the 2nd order gauge invariant Newtonian potential $\Phi$ by
	
		\ba\label{eq:ztop2}
		\zeta=\left(1+\frac{1}{\epsilon}\right)\Phi+\frac{1}{\epsilon H}\dot\Phi-\frac{1}{2a^3\dot\phi^2}\Delta^{-1}&\left(\dot\gamma^N_{ij}\pa_i\pa_j\left[\dot\Phi+H\Phi\right]\right)+\Delta^{-1}\left(\gamma^N_{ij}\pa_i\pa_j\Phi\right)\\&
		-\frac{a^{2}}{6}\Delta^{-1}\left(\dot\gamma^N_{ij}\dot\gamma^N_{ij}\right)-\frac{\Delta^{-1}}{16}\left(\pa_k\gamma^N_{ij}\pa_k\gamma^N_{ij}\right)+O(\Phi^2)\,.
		\ea
		On the other hand, the tensor modes in the uniform-$\phi$ and Newtonian slicing are related by
				\ba\label{eq:gammaCN2}
				\gamma_{ij}^{C}=\gamma_{ij}^{N}+\widehat{TT}_{ij}\,^{ab}\left\{\frac{2}{\dot\phi^2}\left(\dot\Phi+H\Phi\right)\dot\gamma^N_{ab}-\frac{4}{a^2\dot\phi^4}\pa_a\left(\dot\Phi+H\Phi\right)\pa_b\left(\dot\Phi+H\Phi\right)\right\}\,.
				\ea
			
		We have shown that with such transformation we recover the well-known equations of motion for tensor modes at 2nd order sourced by scalar 2nd order perturbations Eq.~\eqref{eq:gwnewton} (see Refs.~\cite{Baumann:2007zm,Ananda:2006af,Alabidi:2012ex}).

	This paper provides a general study of the gauge issues of cosmological perturbations up to 2nd order within the Hamiltonian formalism and constitutes a complementary check to the usual method using the Lie derivatives, e.g. Refs.~\cite{Bruni:1996im,Malik:2008im}. In contrast, the Hamiltonian formalism provides a systematic way to find the canonical definition of the perturbation variables which successfully reduced the system. The new contribution of this work is the reduction of the action without any gauge fixing with the correct definition of scalar and tensor modes that coincide with perturbations in the uniform-$\phi$ slicing. Once the system has been reduced, we have shown that one can easily move from gauge to gauge by means of canonical transformations. We leave for future work which is the definition of tensor modes that we actually observe in the CMB. It would be interesting to see if the 2nd order change in the definition of tensor modes could have any effect on the values of the 3-point functions sources by mixed terms like in Ref.~\cite{Domenech:2017kno}.

	\begin{acknowledgements}
 This work was supported in part by the MEXT KAKENHI Nos. 15H05888 and 15K21733.
Some involved calculations were cross-checked with the Mathematica package xAct (www.xact.es).
\end{acknowledgements}
\appendix
\section{Poisson algebra in the conformal decomposition}
Here we give explicit expressions on the variation of the constraints with respect to the canonical variables. They are useful when computing the poisson algebra. Note that when taking the variation with respect to the $\Upsilon_{ij}$ and $\Pi^{ij}$ one must bear in mind that they are traceless by definition. With that we have that for the Hamiltonian constraint
\ba
\frac{\delta N\ch_N}{\delta {\Upsilon_{ab}}}&=4N\e^{-3\Psi}\Pi^{i(a}\Pi^{b)j}\Upsilon_{ij}-2N\e^{\Psi}D^aD^b\Psi
+2D^{(a}\left(N\e^{\Psi}D^{b)}\Psi\right)
-\Upsilon^{ab}D^{k}\left(N\e^{\Psi}D_{k}\Psi\right)\\&-N\e^{\Psi}D^a\Psi D^b\Psi+\frac{N}{2}\e^\Psi R^{ab}
+\frac{1}{2}\Upsilon^{ab}D_kD^k\left(N\e^\Psi\right)-\frac{1}{2}D^aD^b\left(N\e^\Psi\right)\,,
\ea
\ba
\frac{\delta N\ch_N}{\delta {\Pi^{ab}}}&=N\e^{-3\Psi}\Pi^{ij}\Upsilon_{i(a}\Upsilon_{b)j}\,,
\ea
\ba
\frac{\delta N\ch_N}{\delta {\Psi}}=&-4N\e^{-3\Psi}\left(2\Pi^{ij}\Pi_{ij}-\frac{1}{12}\Pi_\Psi^2\right)+2D_i\left(\e^\Psi D_iN\right)+N\ch\,,
\ea
\ba
\frac{\delta N\ch_N}{\delta {\Pi_\Psi}}&=-\frac{1}{6}N\e^{-3\Psi}\Pi_\Psi\,,
\ea
where we neglected the scalar field $\Phi$ since it's contribution can be added without difficulties. For the momentum constraint we obtain
\ba
\frac{\delta N^i\ch_i}{\delta {\Upsilon_{ab}}}&=2\Pi^{k(a}D_kN^{b)}-\frac{2}{3}\Pi^{ab}D_kN^k-D_k\left(N^k\Pi^{ab}\right)\,,
\ea
\ba
\frac{\delta N^i\ch_i}{\delta {\Pi^{ab}}}&=2D_{(a}N_{b)}-\frac{2}{3}D_kN^k\Upsilon_{ab}\,,
\ea
\ba
\frac{\delta N^i\ch_i}{\delta {\Psi}}&=-D_i\left(N^i\Pi_\Psi\right)\,,
\ea
\ba
\frac{\delta N^i\ch_i}{\delta {\Pi_\Psi}}&=N^iD_i\Psi+\frac{1}{3}D_kN^k\,.
\ea
\section{Poisson algebra for perturbations up to 2nd order \label{app:poissonpert}}
Here we explicitly show the variation of the constraints with respect to the canonical perturbation variables up to 2nd order. Again, we must bear in mind that $Y_{ij}$ and $P^{ij}$ are both traceless. This time, the scalar field $\vp$ has been included. We find for the Hamiltonian constraint that
\ba
\frac{\delta A\ch_{N,1+2}}{\delta {Y_{ab}}}=&-\frac{a}{2}D_{ab}A-\frac{a}{2}D_{ab}\left(A\psi\right)+2a\pa_{(a}A\pa_{b)}\psi-\frac{2a}{3}\delta_{ab}\pa_kA\pa_k\psi
\\&+\frac{a}{2}\pa_k\left(^{1}\Gamma^k_{ab}A\right)-\frac{a}{6}\delta_{ab}\pa_k\left(A\pa_lY_{kl}\right)
+\frac{a}{2}Y_{k(a}\pa_{b)}\pa_kA-\frac{a}{6}\delta_{ab}Y_{kl}\pa_k\pa_l A\,,
\ea
\ba
\frac{\delta A\ch_{N,1+2}}{\delta {P^{ab}}}&=4a^{-3}AP_{ab}\,,
\ea
\ba
\frac{\delta A\ch_{N,1+2}}{\delta {\psi}}=&6a^3VA+2a\Delta A-3A\ch_0+2a\pa_k\left(\psi\pa_kA\right)\\&+A\left(18a^3\psi V+6a^3V_\phi\vp+8a\Delta\psi-2a\pa_k\pa_lY_{kl}\right)
-2a\pa_k\left(Y_{kl}\pa_lA\right)
-3A\ch_1\,,
\ea
\ba
\frac{\delta A\ch_{N,1+2}}{\delta {\pi_\psi}}&=-\frac{A}{6}a^{-3}\pi_\alpha+\frac{A}{2}a^{-3}\left(\psi\pi_\alpha-\frac{\pi_\psi}{6}\right)\,,
\ea
\ba
\frac{\delta A\ch_{N,1+2}}{\delta {\varphi}}&=a^3V_\phi A+a^3 A\left(3\psi V_\phi+V_{\phi\phi}\vp\right)-a\pa_k\left(A\pa_k\vp\right)\,,
\ea
\ba
\frac{\delta A\ch_{N,1+2}}{\delta {\pi_\varphi}}&=a^{-3}A\pi_\phi+a^{-3}A\left(\pi_\vp-3\psi\pi_\phi\right)\,,
\ea
For the momentum constraint we find
\ba
\frac{\delta \pa^i B\ch_{i,1+2}}{\delta {Y_{ab}}}&=-\pa_k\left(P_{ab}\pa_kB\right)\,,
\ea
\ba
\frac{\delta \pa^i B\ch_{i,1+2}}{\delta {P^{ab}}}&=2D_{ij}B+\pa_kY_{ab}\pa_k B\,,
\ea
\ba
\frac{\delta \pa^i B\ch_{i,1+2}}{\delta {\psi}}&=-\pi_\alpha\Delta B-\pa_k\left(\pi_\psi\pa_kB\right)\,,
\ea
\ba
\frac{\delta \pa^i B\ch_{i,1+2}}{\delta {\pi_\psi}}&=\frac{1}{3}\Delta B+\pa_k B\pa_k\psi\,,
\ea
\ba
\frac{\delta \pa^i B\ch_{i,1+2}}{\delta {\vp}}&=-\pi_\phi\Delta B-\pa_k\left(\pi_\vp\pa_kB\right)\,,
\ea
\ba
\frac{\delta \pa^i B\ch_{i,1+2}}{\delta {\pi_\vp}}&=\pa_k\vp\pa_k B\,.
\ea
These are the formulas used in the main body of the paper to compute the gauge transformation of the perturbation variables. One can also explicitly check that these formulas satisfy the poisson algebra which holds up to 3rd order in perturbation.
\section{Perturbation expansion of the variables \label{app:perturbvar}}

We expand into a time dependent background and perturbations as follows. First, we split the conformal degree of freedom and the scalar field as
	\ba
	\Psi=\alpha(t)+\psi(t,\mathbf{x}) \quad {\rm and}\quad \Theta=\phi(t)+\varphi(t,\mathbf{x})\,,
	\ea
	where $\alpha\equiv\ln a$, $a(t)$ is the scale factor of the FLRW expanding universe and $\phi(t)$ is the background value of the inflaton. The respective conjugate momenta are given by
	\ba
	\Pi_\Psi=\pi_{\alpha}(t)+\pi_\psi(t,\mathbf{x})\quad{\rm and}\quad \Pi_\Theta=\pi_\phi(t)+\pi_\varphi(t,\mathbf{x})\,.
	\ea
	Note that in terms of the usual quantities $\pia$ and $\pip$ are respectively given by
	\ba
	\pi_\alpha=-6a^3H \quad{\rm and}\quad \pi_\phi=a^3\dot\phi \,,
	\ea
	where $H\equiv{\dot a}/{a}$.
	We expand the lapse and shift as
	\ba
	N=1+A \quad{\rm and}\quad N_i=a^{-2}\pa_i B\,.
	\ea
	Now we are left with the expansion of the traceless degrees of freedom $\Upsilon_{ij}$. In doing so, we will extract the scalar and tensor degrees contained in $\Upsilon_{ij}$. First of all, we define in general the perturbations as
	\ba
	Y_{ij}\equiv\left[\ln\Upsilon\right]_{ij}\,,
	\ea
	in other words $\Upsilon_{ij}$ is an exponential function of the perturbations. In this way, the condition $\det \Upsilon=1$ is automatically satisfied. One can check that as expected $Y_{ij}$ is traceless, i.e. $\delta^{ij}Y_{ij}=0$. We define the scalar degree so as to match the usual definition,  as
	\ba
	E=\frac{3}{4}\Delta^{-2}\pa_k\pa_lY_{kl}\,,
	\ea
	where $\Delta^{-1}$ is the inverse Laplacian operator, $\Delta\equiv\pa_i\pa^i$ and we assume that it is well defined and the fields vanish at infinity. On the other hand, the \textit{transverse-traceless} component, i.e. the tensor modes, is defined as
	\ba
	\gamma_{ij}=\widehat{TT}_{ij}\,^{ab}Y_{ab}\,,
	\ea
	where $\widehat{TT}_{ij}\,^{ab}$ is the transverse-traceless projector and it is given by
	\ba\label{eq:tt}
	\widehat{TT}_{ij}\,^{ab}=&
	\left(\delta_{i}^{(a}-\pa_i\pa^{(a}\Delta^{-1}\right)\left(\delta_j^{b)}-\pa_j\pa^{b)}\Delta^{-1}\right)-\frac{1}{2}\left(\delta_{ij}-\pa_i\pa_j\Delta^{-1}\right)\left(\delta^{ab}-\pa^a\pa^b\Delta^{-1}\right)
	\ea
	where symmetrization of indexes is normalized.
	With these definitions we find that
	\ba
	Y_{ij}=\gamma_{ij}+2D_{ij}E\,,
	\ea
	where we have defined for convenience the traceless second derivative operator as
	\ba
	D_{ij}\equiv\partial_i\partial_j - \frac{1}{3}\delta_{ij}\Delta\,.
	\ea
	
	On the other hand, the expansion of the corresponding conjugate momenta for $Y_{ij}$ valid up to 4th order is given by
	\ba
	\Pi^{ij}=P^{l(i}\delta_{lk}\Upsilon^{j)k}+O(4)
	\ea
	As before, we define the conjugate momenta of $E$ and $\gamma_{ij}$ respectively as
	\ba
	\pi_E=2\pa_a\pa_bP^{ab}\quad{\rm and}\quad \pi_{ij}=\widehat{TT}^{ij}\,_{ab}P^{ab}\,.
	\ea
	Then the expansion of $P^{ij}$ in terms of $\pi_E$ and $\pi_{ij}$ reads
	\ba
	P^{ij}=\pi_{ij}+\frac{3}{4}D^{ij}\Delta^{-2}\pi_E\,.
	\ea
	These change of variables does not modify the Hamiltonian as it can be seen from the fact that
	\ba
	\Pi^{ij}\dot\Upsilon_{ij}=P^{ij}\dot Y_{ij}+O(4)=\pi_{ij}\dot\gamma_{ij}+\pi_E\dot E+O(4)\,.
	\ea

\section{Perturbation expansion of the Hamiltonian \label{app:perturbhamiltonian}}
In this appendix we present the explicit form of the Hamiltonian expansion without any integration by parts.
We expand the Hamiltonian constraint as follows,
\ba
	\begin{split}
		\ch_N&={{\rm e}^{-3\Psi}}\Pi+{\rm e}^{3\Psi}V(\Theta)+ {\rm e}^{\Psi} J=\sum_{i=0}^3\ch_{N,i}+O(4)\,.
	\end{split}\,,
\ea
where we have defined
\ba
	\Pi\equiv a^{-3}\left(2\Pi^{ij}\Pi_{ij}-\frac{\Pi_\Psi^2}{12}+\frac{\Pi_{\Theta}^2}{2}\right)=\sum_{i=0}^3\Pi_i+O\left(4\right)\,,
\ea
\ba
	V(\Theta)=\sum_{i=0}^3\frac{1}{i!}\frac{\pa^iV}{\pa\Theta^i}\Bigg|_{\phi(t)}\varphi^i+O(4)\,
\ea
and
\ba
	J\equiv & a \left(2\Upsilon^{ij}D_iD_j\Psi+\Upsilon^{ij}D_i \Psi D_j \Psi-\frac{1}{2}R^{(3)}+\frac{1}{2}\Upsilon^{ij} D_i \Theta D_j \Theta\right)=\sum_{i=0}^3J_i+O(4)\,.
\ea
Using that the metric is unit determinant we have a simple expression for the Ricci scalar, which is given by
\ba
R^{(3)}=-\pa_i\pa_j\Upsilon^{ij}-\Gamma^{i}_{kl}\pa_i\Upsilon^{kl}-\Upsilon^{kl}\Gamma^i_{kj}\Gamma^j_{il}=\sum_{i=0}^3R^{(3)}_i+O(4)\,,
\ea
where $\Gamma^{i}_{kl}=\tfrac{1}{2}\Upsilon^{ij}\left(2\pa_{(k}\Upsilon_{l)j}-\pa_j\Upsilon_{kl}\right)$ are the Christoffel symbols.
Identifying term by term in the expansion we have the following terms for $\Pi$:
\ba
	&a^3\Pi_0=\frac{\pi_\phi^2}{2}-\frac{\pi_\alpha^2}{12}\quad{;}\quad a^3\Pi_1=\pi_\phi\pi_\varphi-\frac{1}{6}\pi_\alpha\pi_\psi\\
	&a^3\Pi_2=2P^{ij}P_{ij}+\frac{\pi_\varphi^2}{2}-\frac{\pi_\psi^2}{12}\quad{;}\quad \Pi_3=0\,.
\ea
For the term $J$ we obtain
\ba
	&J_0=0\quad{;}\quad J_1/a=2\Delta\psi-\frac{1}{2}R^{(3)}_1\,,\\
	&J_2/a=-2\pa_i\left(Y_{ij}\pa_j\psi\right)+\pa_k\psi\pa_k\psi+\frac{1}{2}\pa_k\varphi\pa_k\varphi-\frac{1}{2}R_2^{(3)}\,,\\
	&J_3/a=\pa_i\left(Y_{ik}Y_{kj}\pa_j\psi\right)-Y_{ij}\pa_i\psi\pa_j\psi-\frac{1}{2}Y_{ij}\pa_i\varphi\pa_j\varphi-\frac{1}{2}R_3^{(3)}\,.
\ea
Lastly, for the Ricci scalar we have
\ba
	&R^{(3)}_1=\pa_i\pa_kY_{ik}\quad ,\quad
	R_2^{(3)}=-\pa_i\left(Y_{ik}\pa_lY_{lk}\right)+\frac{1}{2}\pa_iY_{ik}\pa_lY_{lk}-\frac{1}{4}\pa_iY_{kl}\pa_iY_{kl},,\\
	&R_3^{(3)}=\frac{1}{6}\pa_i\pa_j\left(Y_{ik}Y_{kl}Y_{lj}\right)+\frac{1}{4}Y_{ij}\pa_jY_{kl}\left(\pa_iY_{kl}-2\pa_{(k}Y_{l)i}\right)\,.
\ea
Then we find
\ba
	&\ch_{N,1}=\Pi_1+V_1+J_1+9\psi V_0\quad ,\quad \ch_{N,2}=\Pi_2+V_2+J_2-3\psi\Pi_1+3\psi V_1+\psi J_1\\
	&\ch_{N,3}=V_3+J_3-3\psi\Pi_2+3\psi V_2+\psi J_2+\frac{9}{2}\psi^2\Pi_1+\frac{9}{2}\psi^2V_1+\frac{1}{2}\psi^2 J_1+{9}\psi^3V_0\,,
\ea
where we already used the background equations of motion $\Pi_0=V_0$.
On the other hand, the momentum constraint is given by
\ba
	\begin{split}
		\ch_i&=\Pi_\Theta D_i\Theta+\Pi_\Psi D_i\Psi-\frac{1}{3}D_i\Pi_\Psi-2\Upsilon_{ij}D_k\Pi^{kj}
		=\ch_{i,1}+\ch_{i,2}+O(3)
	\end{split}
\ea
where
\ba
	\begin{split}
		\ch_{i,1}&=\pi_\phi \pa_i\varphi+\pi_\alpha \pa_i\psi-\frac{1}{3}\pa _i\pi_\psi-2\delta_{ij}\pa_kP^{kj}
	\end{split}
\ea
and
\ba
	\begin{split}
		\ch_{i,2}&=
		\pi_\varphi \pa_i\varphi+\pi_\psi \pa_i\psi+P^{kl}\pa_i Y_{kl}\,,
	\end{split}
\ea
where in the last equality we used the fact that $N^i=\pa_i\beta$, i.e. we neglected the vector modes and therefore some terms are simplified. We finish this appendix by giving the following identities which are heavily used throughout the main body calculations:
\ba
\pa_i\pa_jA\pa_i\pa_jB-\Delta A\Delta B=\left(\Delta\delta_{ij}-\pa_i\pa_j\right)\left[\pa_i A\pa_jB\right]
\ea
and
\ba
\pa_i A\pa_jA\left(\Delta\delta_{ij}-\pa_i\pa_j\right)A=\frac{3}{2}\Delta A\pa_i A\pa_i A-\frac{1}{2}\pa_j\left(\pa_i A\pa_i A\pa_jA\right)\,.
\ea

\newpage

\section{Explicit expressions of the canonical variables \label{app:canonicalvariables}}
Here we present the detailed expression for the canonical momenta of the curvature perturbation which are gauge invariant up to 2nd order. First, we have that Eq.~\eqref{eq:pizeta} is completed by

	\ba\label{eq:pizetaapp}
	\pi_\omega^{GI}&\equiv\pi_\omega-\pa_i\left(\pi_\omega\pa_iE\right)-\frac{\pia}{2}\Delta E\Delta E+\frac{6a^6V}{\pip}\vp\Delta E-\frac{2a^4}{\pip^2}\Delta\left(\vp\pi_2\right)-\frac{2a^4}{\pia\pip}\Delta\left(\vp\pi_1\right)
	-\frac{6a^6V}{\pip^2}\vp\pi_2\\&-\frac{\pia}{2\pip}\vp\pi_1-\frac{2a^4}{\pip}\pa_i\omega\pa_i\vp-\frac{2a^4}{\pip}\gamma_{ij}\pa_i\pa_j\vp+\frac{2a^4}{\pip}\Delta\left(\omega\vp\right)+\frac{18a^6V}{\pip}\omega\vp+\frac{a^4\pia}{6\pip^2}\pa_i\vp\pa_i\vp\\&
	+\frac{3a^6}{2\pip}\left(V_\phi-\frac{\pia}{\pip}V\right)\varphi^2-\left(\frac{a^4\pia}{6\pip^2}+\frac{3a^4}{\pia}\right)\Delta\vp^2-\frac{12a^4}{\pia\pip}\pi_{ij}\pa_i\pa_j\vp
	-\frac{3a^4}{\pia}\pa_k\gamma_{ij}\pa_i\pa_j\pa_kE\\&+\frac{3a^4}{\pia}\left(\pa_i\pa_j-\Delta\delta_{ij}\right)\pa_i\pa_kE\pa_j\pa_kE-3\pia\omega\Delta E
	+\frac{6a^4}{\pip}\vp\Delta\omega+\frac{27}{4\pia}\left(\pa_i\pa_j-\Delta\delta_{ij}\right)\pa_i\Delta^{-2}\pi_E\pa_j\Delta^{-2}\pi_E\\&
	+\frac{4a^4}{\pip}\left(\pa_i\pa_j-\Delta\delta_{ij}\right)\pa_i E\pa_j \vp-\frac{a^4\pia}{\pip^2}\vp\Delta\Xi
	-\frac{24a^4}{\pia}\left(\pa_i\pa_j-\Delta\delta_{ij}\right)\pa_i E\pa_j \omega\\&+\frac{12a^4}{\pia}\pa_i\left(\pa_i\omega\Delta E\right)-\frac{2a^4}{\pip}\pa_i\left(\Delta E\pa_i\vp\right)\,.
	\ea
	For simplicity we have introduced the gauge invariant variables at 1st order
	\ba\label{eq:xi2}
	\pi_2\equiv\pi_\vp-3\pip\psi-\frac{a^6V_\phi}{\pip}\vp-\frac{\pia}{2}\vp=\pi_{\dphi}+\frac{\pia}{6\pip}\pi_\omega
\quad {\rm and}\quad
	\Xi\equiv\vp-\frac{3\pip}{2a^4}\Delta^{-2}\pi_E=\frac{\pip}{2a^4}\Delta^{-1}\pi_\omega+6\frac{\pip}{\pia}\omega\,,
	\ea
	where the equalities hold once the 1st order constraints are solved. As we will later see, the quantity $\Xi$ in Eq.~\eqref{eq:xi}, will be useful in the Newtonian slicing, Sec.~\ref{sec:newtonian}, as it gives the difference between the slicings with $\vp$ and $\pi_E$. The full expression of Eq.~\eqref{eq:pizeta2} is
		\ba\label{eq:pizeta2app}
		\pi_\zeta&\equiv\pi_\omega-\pa_i\left(\pi_\omega\pa_iE\right)-\frac{\pia}{2}\Delta E\Delta E+\frac{6a^6V}{\pip}\vp\Delta E+\frac{18a^6V}{\pip}\omega\vp-3\pia\omega\Delta E
		-\frac{2a^4}{\pip}\pa_i\omega\pa_i\vp\\&-\frac{2a^4}{\pip}\gamma_{ij}\pa_i\pa_j\vp+\frac{2a^4}{\pip}\Delta\left(\omega\vp\right)+\frac{a^4\pia}{6\pip^2}\pa_i\vp\pa_i\vp
		+\frac{3a^6}{2\pip}\left(V_\phi-\frac{\pia}{\pip}V\right)\varphi^2-\left(\frac{a^4\pia}{6\pip^2}+\frac{3a^4}{\pia}\right)\Delta\vp^2\\&-\frac{12a^4}{\pia\pip}\pi_{ij}\pa_i\pa_j\vp
		-\frac{3a^4}{\pia}\pa_k\gamma_{ij}\pa_i\pa_j\pa_kE+\frac{3a^4}{\pia}\left(\pa_i\pa_j-\Delta\delta_{ij}\right)\pa_i\pa_kE\pa_j\pa_kE+\frac{9}{2}\pia\zeta^2\\&
		+\frac{4a^4}{\pip}\left(\pa_i\pa_j-\Delta\delta_{ij}\right)\pa_i E\pa_j \vp
		-\frac{24a^4}{\pia}\left(\pa_i\pa_j-\Delta\delta_{ij}\right)\pa_i E\pa_j \omega+\frac{12a^4}{\pia}\pa_i\left(\pa_i\omega\Delta E\right)\\&-\frac{2a^4}{\pip}\pa_i\left(\Delta E\pa_i\vp\right)
		+\frac{3a^4}{\pia\pip}\left(\Delta\delta_{ij}-\pa_i\pa_j\right)\left[\pa_i\vp\pa_j\Delta^{-1}\pi_\omega+\frac{12a^4}{\pia}\pa_i\vp\pa_j\omega-\frac{a^4}{\pip}\pa_i\vp\pa_j\vp\right]\,.
		\ea
\newpage
\section{Change in the Hamiltonian at 3rd order\label{app:changelagrangian}}
Since change in the 3rd order Hamiltonian is quite involved, we present the full expression in this appendix. After the canonical transformation Eqs.~\eqref{eq:zeta2}-\eqref{eq:piij2}  we find
\ba
\delta\ch&=-\dt{\frac{2}{\pip}}\pi_{ij}\pi_{ij}\vp-\dt{\frac{a^4}{8\pip}}\vp\pa_k{\gamma_{ij}}\pa_k{\gamma_{ij}}+\dt{\frac{3a^4}{\pia}}\zeta\pa_k\gamma_{ij}\pa_i\pa_j\pa_k E\\&
-\dt{\frac{1}{\pip}}\pi_{ij}\pa_i\Delta^{-1}\pi_\zeta\pa_j\vp+\dt{\frac{a^4\pia}{6\pip^2}}\gamma_{ij}\pa_i\vp\pa_j\vp-\dt{\frac{2a^4}{\pip}}\gamma_{ij}\pa_i\vp\pa_j\zeta
\\&+\dt{\frac{a^4}{\pip^2}}\pi_{ij}\pa_i\vp\pa_j\vp-\dt{\frac{12a^4}{\pip\pia}}\pi_{ij}\pa_i\zeta\pa_j\vp-\dt{\frac{a^4}{2\pip}}\vp\pa_k\gamma_{\ij}\pa_i\pa_j\pa_k E
\\&+\dt{\frac{\pia}{2}}\zeta \Delta E \Delta E+\dt{\frac{3a^4}{\pia}}\pa_i\pa_kE\pa_j\pa_kE\left[\Delta\delta_{ij}-\pa_i\pa_j\right]\zeta-\dt{\frac{a^6V}{\pip}}\vp\Delta E\Delta E
\\&-\dt{\frac{a^4}{2\pip}}\pa_i\pa_kE\pa_j\pa_kE\left[\Delta\delta_{ij}-\pa_i\pa_j\right]\vp+\dt{\frac{3\pia}{2}}\zeta^2\Delta E+\dt{\frac{6a^4}{\pia}}\Delta E\pa_i\zeta\pa_i\zeta
\\&-\dt{\frac{6a^6V}{\pip}}\zeta\vp\Delta E-\dt{\frac{2a^4}{\pip}}\Delta E\pa_i\zeta\pa_i\vp+\dt{\frac{a^4\pia}{6\pip^2}}\Delta E\pa_i\vp\pa_i\vp
\\&-\dt{\frac{a^6}{2\pip}\left(V_\phi-\frac{\pia}{\pip}V\right)}\vp^2\Delta E+\dt{\frac{a^4}{4\pip^2}}\pa_i\vp\pa_j\vp\left[\delta_{ij}-\pa_i\pa_j\Delta^{-1}\right]\pi_\zeta
\\&+\dt{\frac{\pia}{18}}\Delta E \Delta E \Delta E-\dt{\frac{3a^6}{2\pip}\left(V_\phi-\frac{\pia}{\pip}V\right)}\zeta\vp^2
-\dt{\frac{a^6}{2\pip^2}\left(V-\frac{\pia}{6\pip}V_\phi\right)}\pi_\zeta\vp^2
\\&+\dt{\frac{3a^8}{\pia\pip^2}}\pa_i\vp\pa_j\vp\left[\Delta\delta_{ij}-\pa_i\pa_j\right]\zeta+\dt{\frac{3a^4}{\pia}}\vp^2\Delta\zeta
+\dt{\frac{a^4\pia}{6\pip^2}}\vp^2\Delta\zeta-\zeta\pa_i\vp\pa_i\vp
\\&-\dt{\frac{a^4}{4\pip}\left(1+\frac{\pia^2}{18\pip^2}\right)}\vp^2\Delta\vp-\dt{\frac{a^8}{4\pip^3}}\pa_i\vp\pa_i\vp\Delta\vp
\\&-\dt{\frac{a^6}{3\pip}\left(\frac{1}{2}V_{\phi\phi}-\frac{3a^6}{\pip^2}\left(V^2-\frac{1}{6}V_{\phi}^2\right)-\frac{\pia}{\pip}\left(V_\phi-\frac{\pia}{2\pip}V\right)\right)}{\vp^3}-\dt{\frac{\pia^2}{72\pip^3}}\pi_\zeta^2\vp\\&
-\dt{\frac{9a^6V}{\pip}}\zeta^2\vp-\dt{\frac{a^4}{\pip}}\zeta^2\Delta\vp+\dt{\frac{a^4}{\pip}}\vp\pa_i\zeta\pa_i\zeta+\dt{\frac{18a^8}{\pia^2\pip}}\pa_i\zeta\pa_j\zeta\left[\pa_i\pa_j-\Delta\delta_{ij}\right]\vp\\&
-\dt{\frac{3a^4}{\pia\pip}}\pa_i\vp\pa_j\zeta\left(\Delta\delta_{ij}-\pa_i\pa_j\right)\Delta^{-1}\pi_\zeta-\dt{\frac{1}{8\pip}}\pa_i\Delta^{-1}\pi_\zeta\pa_j\Delta^{-1}\pi_\zeta\left(\Delta\delta_{ij}-\pa_i\pa_j\right)\vp\,.
\ea
\newpage

\section{Change in the Hamiltonian from uniform-\texorpdfstring{$\phi$}{phi} to Newtonian slicing\label{app:changelagrangian2}}
At 2nd order in the action we have that
\ba
\delta \ch=-\dt{\frac{6a^4}{\pia}\left(1+\frac{\pia^2}{18\pip^2}\right)}\Phi\Delta\Phi+\frac{12a^4}{\pia}\dt{\frac{\pia^2}{18\pip^2}}\Phi\Delta\Phi+\frac{a^4}{\pia}\dt{\frac{\pia}{a^4}}\pi_\Phi\Phi\,
\ea
and at 3rd order we find
\ba
\delta \ch&=\frac{\pia}{\pip}\dt{\frac{a^4}{8\pia}}\Xi\pa_k\gamma_{ij}\pa_k\gamma_{ij}+\frac{\pia}{\pip}\dt{\frac{2a^4}{\pia}}\gamma_{ij}\pa_i\Phi\pa_j\Xi\\&+\frac{2\pia}{\pip}\dt{\frac{1}{\pia}}\Xi\pi_{ij}\pi_{ij}
+\dt{\frac{12a^4}{\pia}}\left(a^{-4}\Phi\pi_{ij}\pi_{ij}+\frac{1}{16}\Phi\pa_k\gamma_{ij}\pa_k\gamma_{ij}\right)
\\&
+\left(\frac{\pia}{\pip}\dt{\frac{a^4}{6\pip}}-\frac{a^4}{6\pip}\dt{\frac{\pia}{\pip}}\right)\gamma_{ij}\pa_i\Xi\pa_j\Xi+\dt{\frac{12a^4}{\pia}}\gamma_{ij}\pa_i\Phi\pa_j\Phi\\&
+\left(\dt{\frac{a^4}{\pip^2}}-\frac{2a^4}{\pia\pip}\dt{\frac{\pia}{\pip}}\right)\pi_{ij}\pa_i\Xi\pa_j\Xi+\frac{1}{\pip}\dt{\frac{12a^4}{\pia}}\pi_{ij}\pa_i\Phi\pa_j\Xi\,.
\ea

\end{document}